\RequirePackage{fix-cm}
\documentclass[twocolumn]{svjour3}          

\smartqed  
\usepackage{cite}
\usepackage{amsmath,amssymb,amsfonts}
\usepackage{amssymb}
\usepackage{amsfonts} 
\usepackage{algorithmic}
\usepackage{graphicx}
\usepackage{textcomp}
\usepackage{xcolor}
\usepackage{enumerate}
\usepackage{tabularx}
\usepackage{multirow}
\usepackage{booktabs}
\usepackage{hyperref}
\usepackage{balance}
\usepackage{paralist}
\usepackage[authoryear, round]{natbib}
\usepackage[braket, qm]{qcircuit}

\usepackage{float}

\usepackage{braket}

\usepackage{xcolor}
\usepackage{tikzscale}
\usepackage{tikz}
\usepackage[title]{appendix}
\usetikzlibrary{shapes,arrows,decorations.pathreplacing,calc,arrows.meta,automata,shapes}
\usetikzlibrary{arrows,automata}
\usetikzlibrary{decorations.pathmorphing}
\usetikzlibrary{angles,quotes}
\usetikzlibrary{positioning}
\usetikzlibrary{fit}
\usetikzlibrary{mindmap, trees}
\usetikzlibrary{backgrounds}
\usetikzlibrary{intersections}
\usetikzlibrary{hobby,decorations.markings}
\usetikzlibrary{snakes}
\usepackage{pgfplots}
\usepackage{diagbox}
\usepgfplotslibrary{groupplots,units}
\usepackage{pgfplotstable}
\usepackage{caption}
\usepackage{subcaption}
\usepackage{tikz-qtree}
  \pgfdeclarelayer{background1}
  \pgfdeclarelayer{background2}
  \pgfsetlayers{background2,background1,main}

\journalname{Quantum Machine Intelligence}
\begin{document}

\title{Evaluation of Parameterized Quantum Circuits: on the 
relation between classification accuracy, expressibility and entangling capability
}

\titlerunning{Evaluation of Parameterized Quantum Circuits}        

\author{
Thomas Hubregtsen~$^{1,2,3}$ 
        \and Josef Pichlmeier~$^{3}$ 
        \and Patrick Stecher~$^{4}$ 
        \and Koen Bertels~$^{3}$ 
}


\institute{
Corresponding author: Thomas Hubregtsen\\
\email{thomas.hubregtsen@bwmgroup.com}\\
$^{1}$BMW Group Research, New Technologies, Innovations, Garching bei M\"unchen, Germany
\\$^{2}$Dahlem Center for Complex Quantum Systems, Freie Universitaet Berlin, Berlin, Germany
\\$^{3}$Delft University of Technology, Delft, Netherlands
\\$^{4}$Technical University of Munich, Department of Informatics,
Garching bei M\"unchen, Germany
\\
}
\date{Received: August 2020 / Accepted: [draft]}

\maketitle

\begin{abstract}
An active area of investigation in the search for quantum advantage is Quantum Machine Learning. Quantum Machine Learning, and Parameterized Quantum Circuits in a hybrid quantum-classical setup in particular, could bring advancements in accuracy by utilizing the high dimensionality of the Hilbert space as feature space. But is the ability of a quantum circuit to uniformly address the Hilbert space a good indicator of classification accuracy? In our work, we use methods and quantifications from prior art to perform a numerical study in order to evaluate the level of correlation. We find a strong correlation between the ability of the circuit to uniformly address the Hilbert space and the achieved classification accuracy for circuits that entail a single embedding layer followed by 1 or 2 circuit designs. This is based on our study encompassing 19 circuits in both 1 and 2 layer configuration, evaluated on 9 datasets of increasing difficulty. Future work will evaluate if this holds for different circuit designs. 
\end{abstract}

\keywords{Quantum Neural Networks \and Parameterized Quantum Circuits \and Expressibility \and Quantum Machine Learning\and Quantum Computing \and Entangling capability}

\section{\bf{Introduction}} \label{sec:introduction}
Quantum computing has seen a steady growth in interest ever since the quantum supremacy experiment\citet{th_google_supr}. The search for quantum advantage, quantum supremacy for practical applications, is an active area of research\citep{th_advantage}. One potential domain of applications is Machine Learning\citep{th_advantage2}. Here, quantum computing is said to potentially bring speedups and improvements in accuracy. One line of reasoning to assume an improvement in accuracy is as follows. A classical Neural Network takes the input data and maps it into a higher dimensional feature space. It then, using a combination of learnable linear transformations and static non-linear transformations, maps the data between various higher dimensional feature spaces. This mapping is repeated until the data points are positioned in such a way that a hyperplane can separate data that belongs to different output classes. Given that qubits, the smallest units of information in a quantum computer, can span a larger space due to their quantum mechanical properties, one would expect that with the same resources, data could be mapped between higher-dimensional and larger feature spaces. This would allow a more accurate separation of the data. Or, as a trade-off, one would require fewer resources to address the same space and maintain the same level of accuracy. 

Recently, a framework compatible with shallow depth circuits for Noisy Intermediate Scale Quantum\citep{th_preskill_nisq} systems has been developed, named the hybrid quantum-classical framework. In this framework, the quantum machine leverages \emph{Parameterized Quantum Circuits} (\emph{PQCs}) in order to make predictions and approximations, while the classical machine is used to update the parameters of the circuit\citep{th_mcclean_hqc}. Example algorithms are the Variational Quantum Eigensolver\citep{th_perruzo_vqe} and the Quantum Approximate Optimisation Algorithm\citep{th_farhi_qaoa}. Variational Quantum Circuits can also be used for Machine Learning problems, and bear resemblance to the structure of classical Neural Networks\citep{th_schuld_2018}. 

Just as in classical Neural Networks, many circuit architectures exist. An active area of research focuses on determining the power and capabilities of these circuits\citep{th_brian, th_sukin_2019, th_schuld_2020_de}. Following the previous line of reasoning of mapping data in larger feature spaces, one way of quantifying the power of a PQC is by quantifying the ability of a PQC to uniformly reach the full Hilbert space\citep{th_sukin_2019}. Our work investigates these definitions by performing a numerical analysis on the correlation between such descriptors and classification accuracy in order to guide the choice and design of PQCs, as well as provide insight into potential limitations.

The remainder of this paper is structured as follows. In Section 2, we will describe our approach. Section 3 will outline our experimental setup and design choices. Our results are presented in Section 4 and discussed in Section 5. We will end this paper with the conclusion in Section 6. Raw experimental data can be found in the Appendices. 



\section{\bf{Approach}} \label{sec:approach}

We will use prior art for descriptors that quantify the ability of a PQC to explore the Hilbert space. We will perform numerical simulations on various custom datasets of increasing difficult to quantify classification performance of these circuits. The dataset will be split in a train, test, and validation set. The train set will be used for training the PQC and the test set for a hyperparameter search. We will retrain the PQC for the best hyperparameters and perform the final classification on the validation set in order to create the data points that we will use in our search for correlation. We will repeat the experiments on the validation set to make sure the outcome is consistent. We will perform a statistical analysis to evaluate a potential relation between the descriptors and the classification accuracy of the various circuits. 
\section{\bf{Evaluation}} \label{sec:evaluation}


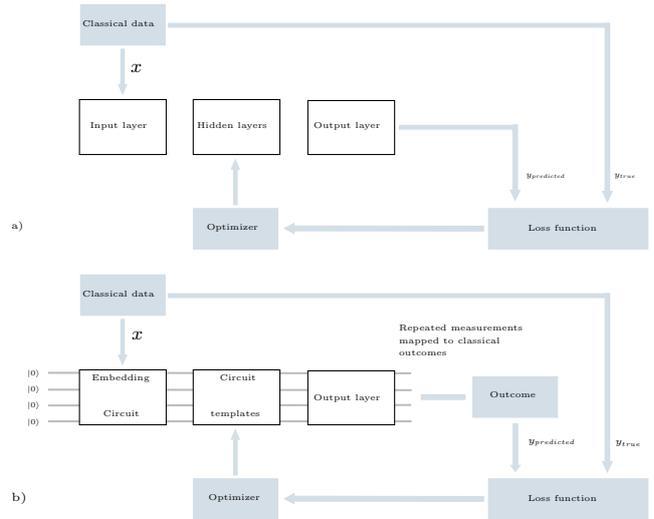
\begin{figure}
\begin{tikzpicture}[x=0.75pt,y=0.75pt, scale=0.83, every node/.style={scale=0.83}, xscale = 0.57, yscale = -0.6]

    \draw [color={rgb, 255:red, 187; green, 187; blue, 187 }  ,draw opacity=1 ][fill={rgb, 255:red, 209; green, 209; blue, 209 }  ,fill opacity=1 ][line width=0.75]    (55,465) -- (438.5,465) ;
    \draw [color={rgb, 255:red, 187; green, 187; blue, 187 }  ,draw opacity=1 ][fill={rgb, 255:red, 209; green, 209; blue, 209 }  ,fill opacity=1 ][line width=0.75]    (55,449) -- (438,448.5) ;
    \draw [color={rgb, 255:red, 187; green, 187; blue, 187 }  ,draw opacity=1 ][fill={rgb, 255:red, 209; green, 209; blue, 209 }  ,fill opacity=1 ][line width=0.75]    (55,433) -- (438.5,433.5) ;
    \draw [color={rgb, 255:red, 187; green, 187; blue, 187 }  ,draw opacity=1 ][fill={rgb, 255:red, 209; green, 209; blue, 209 }  ,fill opacity=1 ][line width=0.75]    (55,416) -- (438,416.5) ;
    \draw  [color={rgb, 255:red, 255; green, 255; blue, 255 }  ,draw opacity=1 ][fill={rgb, 255:red, 211; green, 222; blue, 230 }  ,fill opacity=1 ] (88,43) -- (179.5,43) -- (179.5,86) -- (88,86) -- cycle ;
    \draw  [color={rgb, 255:red, 255; green, 255; blue, 255 }  ,draw opacity=1 ][fill={rgb, 255:red, 211; green, 222; blue, 230 }  ,fill opacity=1 ] (136.5,88) -- (136.5,124) -- (139.5,124) -- (133.5,137.37) -- (127.5,124) -- (130.5,124) -- (130.5,88) -- cycle ;
    \draw  [color={rgb, 255:red, 0; green, 0; blue, 0 }  ,draw opacity=1 ][fill={rgb, 255:red, 255; green, 255; blue, 255 }  ,fill opacity=1 ] (88,140) -- (179.5,140) -- (179.5,195.5) -- (88,195.5) -- cycle ;
    \draw  [color={rgb, 255:red, 255; green, 255; blue, 255 }  ,draw opacity=1 ][fill={rgb, 255:red, 211; green, 222; blue, 230 }  ,fill opacity=1 ] (249.5,247.37) -- (249.5,211.37) -- (246.5,211.37) -- (252.5,198) -- (258.5,211.37) -- (255.5,211.37) -- (255.5,247.37) -- cycle ;
    \draw  [color={rgb, 255:red, 0; green, 0; blue, 0 }  ,draw opacity=1 ][fill={rgb, 255:red, 255; green, 255; blue, 255 }  ,fill opacity=1 ] (208,140) -- (299.5,140) -- (299.5,195.5) -- (208,195.5) -- cycle ;
    \draw  [color={rgb, 255:red, 0; green, 0; blue, 0 }  ,draw opacity=1 ][fill={rgb, 255:red, 255; green, 255; blue, 255 }  ,fill opacity=1 ] (329,141) -- (420.5,141) -- (420.5,195.5) -- (329,195.5) -- cycle ;
    \draw  [color={rgb, 255:red, 255; green, 255; blue, 255 }  ,draw opacity=1 ][fill={rgb, 255:red, 211; green, 222; blue, 230 }  ,fill opacity=1 ] (207,249) -- (298.5,249) -- (298.5,292) -- (207,292) -- cycle ;
    \draw  [color={rgb, 255:red, 255; green, 255; blue, 255 }  ,draw opacity=1 ][fill={rgb, 255:red, 211; green, 222; blue, 230 }  ,fill opacity=1 ] (180.5,62) -- (647.5,62) -- (647.5,68) -- (180.5,68) -- cycle ;
    \draw  [draw opacity=0][fill={rgb, 255:red, 211; green, 222; blue, 230 }  ,fill opacity=1 ] (648,62.5) -- (648,232.19) -- (651,232.19) -- (645,244.5) -- (639,232.19) -- (642,232.19) -- (642,62.5) -- cycle ;
    \draw  [color={rgb, 255:red, 255; green, 255; blue, 255 }  ,draw opacity=1 ][fill={rgb, 255:red, 211; green, 222; blue, 230 }  ,fill opacity=1 ] (518.5,249) -- (689,249) -- (689,292) -- (518.5,292) -- cycle ;
    \draw  [color={rgb, 255:red, 255; green, 255; blue, 255 }  ,draw opacity=1 ][fill={rgb, 255:red, 211; green, 222; blue, 230 }  ,fill opacity=1 ] (424.5,164.96) -- (550.75,164.96) -- (550.75,171) -- (424.5,171) -- cycle ;
    \draw  [draw opacity=0][fill={rgb, 255:red, 211; green, 222; blue, 230 }  ,fill opacity=1 ][line width=0.75]  (550.75,164.96) -- (550.75,231.03) -- (553.75,231.03) -- (547.75,242.96) -- (541.75,231.03) -- (544.75,231.03) -- (544.75,164.96) -- cycle ;
    \draw  [draw opacity=0][fill={rgb, 255:red, 211; green, 222; blue, 230 }  ,fill opacity=1 ] (514.99,273.4) -- (315,273.15) -- (315,276.15) -- (303.25,270.14) -- (315.02,264.15) -- (315.01,267.15) -- (515,267.4) -- cycle ;
    \draw  [color={rgb, 255:red, 255; green, 255; blue, 255 }  ,draw opacity=1 ][fill={rgb, 255:red, 211; green, 222; blue, 230 }  ,fill opacity=1 ] (88,316) -- (179.5,316) -- (179.5,359) -- (88,359) -- cycle ;
    \draw  [color={rgb, 255:red, 255; green, 255; blue, 255 }  ,draw opacity=1 ][fill={rgb, 255:red, 211; green, 222; blue, 230 }  ,fill opacity=1 ] (136.5,361) -- (136.5,397) -- (139.5,397) -- (133.5,410.37) -- (127.5,397) -- (130.5,397) -- (130.5,361) -- cycle ;
    \draw  [color={rgb, 255:red, 0; green, 0; blue, 0 }  ,draw opacity=1 ][fill={rgb, 255:red, 255; green, 255; blue, 255 }  ,fill opacity=1 ] (88,413) -- (179.5,413) -- (179.5,468.5) -- (88,468.5) -- cycle ;
    \draw  [color={rgb, 255:red, 255; green, 255; blue, 255 }  ,draw opacity=1 ][fill={rgb, 255:red, 211; green, 222; blue, 230 }  ,fill opacity=1 ] (249.5,520.37) -- (249.5,484.37) -- (246.5,484.37) -- (252.5,471) -- (258.5,484.37) -- (255.5,484.37) -- (255.5,520.37) -- cycle ;
    \draw  [color={rgb, 255:red, 0; green, 0; blue, 0 }  ,draw opacity=1 ][fill={rgb, 255:red, 255; green, 255; blue, 255 }  ,fill opacity=1 ] (208,413) -- (299.5,413) -- (299.5,468.5) -- (208,468.5) -- cycle ;
    \draw  [color={rgb, 255:red, 0; green, 0; blue, 0 }  ,draw opacity=1 ][fill={rgb, 255:red, 255; green, 255; blue, 255 }  ,fill opacity=1 ] (329,414) -- (420.5,414) -- (420.5,468.5) -- (329,468.5) -- cycle ;
    \draw  [color={rgb, 255:red, 255; green, 255; blue, 255 }  ,draw opacity=1 ][fill={rgb, 255:red, 211; green, 222; blue, 230 }  ,fill opacity=1 ] (207,522) -- (298.5,522) -- (298.5,565) -- (207,565) -- cycle ;
    \draw  [color={rgb, 255:red, 255; green, 255; blue, 255 }  ,draw opacity=1 ][fill={rgb, 255:red, 211; green, 222; blue, 230 }  ,fill opacity=1 ] (180.5,335) -- (647.5,335) -- (647.5,341) -- (180.5,341) -- cycle ;
    \draw  [draw opacity=0][fill={rgb, 255:red, 211; green, 222; blue, 230 }  ,fill opacity=1 ] (648,335.5) -- (648,505.19) -- (651,505.19) -- (645,517.5) -- (639,505.19) -- (642,505.19) -- (642,335.5) -- cycle ;
    \draw  [color={rgb, 255:red, 255; green, 255; blue, 255 }  ,draw opacity=1 ][fill={rgb, 255:red, 211; green, 222; blue, 230 }  ,fill opacity=1 ] (518.5,522) -- (689,522) -- (689,565) -- (518.5,565) -- cycle ;
    \draw  [color={rgb, 255:red, 255; green, 255; blue, 255 }  ,draw opacity=1 ][fill={rgb, 255:red, 211; green, 222; blue, 230 }  ,fill opacity=1 ] (447.5,437) -- (496,437) -- (496,444) -- (447.5,444) -- cycle ;
    \draw  [draw opacity=0][fill={rgb, 255:red, 211; green, 222; blue, 230 }  ,fill opacity=1 ][line width=0.75]  (551,467) -- (551,508.93) -- (554,508.93) -- (548,516.5) -- (542,508.93) -- (545,508.93) -- (545,467) -- cycle ;
    \draw  [draw opacity=0][fill={rgb, 255:red, 211; green, 222; blue, 230 }  ,fill opacity=1 ] (514.99,546.4) -- (315,546.15) -- (315,549.15) -- (303.25,543.14) -- (315.02,537.15) -- (315.01,540.15) -- (515,540.4) -- cycle ;
    \draw  [color={rgb, 255:red, 255; green, 255; blue, 255 }  ,draw opacity=1 ][fill={rgb, 255:red, 211; green, 222; blue, 230 }  ,fill opacity=1 ] (502,419) -- (593.5,419) -- (593.5,462) -- (502,462) -- cycle ;
    
    \draw (89,58) node [anchor=north west][inner sep=0.7pt]  [font=\fontsize{0.43em}{0.74em}\selectfont] [align=left] {{Classical data}};
    \draw (97,161) node [anchor=north west][inner sep=0.75pt]  [font=\fontsize{0.43em}{0.64em}\selectfont] [align=left] {Input layer};
    \draw (210,161) node [anchor=north west][inner sep=0.75pt]  [font=\fontsize{0.43em}{0.64em}\selectfont] [align=left] {Hidden layers};
    \draw (333,161) node [anchor=north west][inner sep=0.75pt]  [font=\fontsize{0.43em}{0.64em}\selectfont] [align=left] {Output layer};
    \draw (220,264) node [anchor=north west][inner sep=0.75pt]  [font=\fontsize{0.43em}{0.64em}\selectfont] [align=left] {Optimizer};
    \draw (559,264) node [anchor=north west][inner sep=0.75pt]  [font=\fontsize{0.43em}{0.64em}\selectfont] [align=left] {Loss function};
    \draw (557,212.4) node [scale=1.0] [anchor=north west][inner sep=0.75pt]  [font=\fontsize{0.39em}{0.56em}\selectfont]  {$y_{predicted}$};
    \draw (650,212.4) node [scale=1.0] [anchor=north west][inner sep=0.75pt]  [font=\fontsize{0.39em}{0.56em}\selectfont]  {$y_{true}$};
    \draw (14,261) node [anchor=north west][inner sep=0.75pt]  [font=\fontsize{0.5em}{0.72em}\selectfont] [align=left] {a)};
    \draw (89,331) node [anchor=north west][inner sep=0.75pt]  [font=\fontsize{0.43em}{0.64em}\selectfont] [align=left] {{Classical data}};
    \draw (99,416) node [anchor=north west][inner sep=0.75pt]  [font=\fontsize{0.43em}{0.64em}\selectfont] [align=left] {Embedding};
    \draw (333,436) node [anchor=north west][inner sep=0.75pt]  [font=\fontsize{0.43em}{0.64em}\selectfont] [align=left] {Output layer};
    \draw (222,537) node [anchor=north west][inner sep=0.75pt]  [font=\fontsize{0.43em}{0.64em}\selectfont] [align=left] {Optimizer};
    \draw (559,537) node [anchor=north west][inner sep=0.75pt]  [font=\fontsize{0.43em}{0.64em}\selectfont] [align=left] {Loss function};
    \draw (559,480.4) node [scale=1.0] [anchor=north west][inner sep=0.75pt]  [font=\fontsize{0.47em}{0.56em}\selectfont]  {$y_{predicted}$};
    \draw (651,482.4) node [scale=1.0] [anchor=north west][inner sep=0.75pt]  [font=\fontsize{0.47em}{0.56em}\selectfont]  {$y_{true}$};
    \draw (14,534) node [anchor=north west][inner sep=0.75pt]  [font=\fontsize{0.6em}{0.72em}\selectfont] [align=left] {b)};
    \draw (140,102.4) node [anchor=north west][inner sep=0.75pt]  [font=\small]  {$x$};
    \draw (141,372.4) node [anchor=north west][inner sep=0.75pt]  [font=\small]  {$x$};
    \draw (111,452) node [anchor=north west][inner sep=0.75pt]  [font=\fontsize{0.43em}{0.64em}\selectfont] [align=left] {Circuit};
    \draw (234,416) node [anchor=north west][inner sep=0.75pt]  [font=\fontsize{0.43em}{0.64em}\selectfont] [align=left] {Circuit};
    \draw (224,452) node [anchor=north west][inner sep=0.75pt]  [font=\fontsize{0.43em}{0.64em}\selectfont] [align=left] {templates};
    \draw (31,410.9) node [scale=0.8] [anchor=north west][inner sep=0.75pt]  [font=\fontsize{0.47em}{0.56em}\selectfont]  {$\ket{0}$};
    \draw (31,426.9) node [scale=0.8] [anchor=north west][inner sep=0.75pt]  [font=\fontsize{0.47em}{0.56em}\selectfont]  {$\ket{0}$};
    \draw (31,443.9) node [scale=0.8] [anchor=north west][inner sep=0.75pt]  [font=\fontsize{0.47em}{0.56em}\selectfont]  {$\ket{0}$};
    \draw (31,460.9) node [scale=0.8] [anchor=north west][inner sep=0.75pt]  [font=\fontsize{0.47em}{0.56em}\selectfont]  {$\ket{0}$};
    \draw (519,433) node [anchor=north west][inner sep=0.75pt]  [font=\fontsize{0.43em}{0.64em}\selectfont] [align=left] {Outcome};
    \draw (423,365) node [anchor=north west][inner sep=0.75pt]  [font=\fontsize{0.43em}{0.64em}\selectfont] [align=left] {Repeated measurements\\mapped to classical \\outcomes};
\end{tikzpicture}
\caption{The structure of a classical neural network (a) and the structure of a quantum neural network (b)}
\label{fig:nn_structure}  
\end{figure}

Variational Quantum Circuits applied for Machine Learning problems bear a resemblance to classical \emph{Neural Networks}(\emph{NN}), which is why they are also sometimes referred to as \emph{Quantum Neural Networks}(\emph{QNNs})\citep{th_Farhi_2018}. They can be used for classification and regression tasks in a supervised learning approach, or as generative models in an unsupervised setup. We will solely focus on supervised prediction. Both NNs and QNNs, shown in Figure \ref{fig:nn_structure}, embed the data into a higher dimensional representation using an input layer. The input layer of a QNN is called an \emph{embedding circuit}. The hidden layers of a QNN are referred to as \emph{circuit templates}, which together constitute a PQC. The variables in these layers, called \emph{weights} in NNs and \emph{parameters} or \emph{variational angles} in QNNs, are initialized along a Gaussian distribution before training starts. The most notable difference between a NN and a QNN is in the way the output is handled. A NN uses an output layer to directly generate a distribution over the possible output classes using only a single run. A QNN needs to be executed several times, referred to as \emph{shots}, before a histogram over the output states can be generated, which still needs to be mapped to the output classes. The predicted output class is compared to the true output class, denoted as $y_{predicted}$ and $y_{true}$ respectively, and the error is quantified by the loss function. This quantification of the error is used to guide the optimizer to adjust the parameters in an iterative process till convergence in the loss is reached. The measure of \emph{classification accuracy} (\emph{Acc}), or simply \emph{accuracy}, is the number of correctly classified samples over the total number of samples. In the remainder of this section, we will first discuss the descriptors of PQCs that we will investigate, before going into more depth on the experimental setup that we use to investigate the correlation between the descriptors and the classification accuracy. 

\subsection{Descriptors of PQCs} 
Describing the performance of a PQC by the ability of the circuit to uniformly address the Hilbert space has been suggested by \citet{th_sukin_2019}. In their theoretical approach, \citet{th_sukin_2019} propose to quantify this by comparing the true distribution of fidelities corresponding to the PQC, to the distribution of fidelities from the ensemble of Haar random states. In practice, they propose to approximate the distribution of fidelities, the overlap of states defined $F = |\braket{\psi_{\theta}|\psi_{\phi}}|^2$, of the PQC. They do this by repeatedly sampling two sets of variational angles, simulate their corresponding states and take the fidelity of the two resulting states to build up a sample distribution $\hat{P}_{PQC}(F; \Theta)$. The ensemble of Haar random states can be calculated analytically: $P_{Haar} = (N-1)(1-F)^{N-2}$, where $N$ is the dimension of the Hilbert space\citep{th_haar}. The measure of \emph{expressibility}, (\emph{Expr}) is then calculated by taking the Kullback-Leibner divergence (KL divergence) between the estimated fidelity distribution and that of the Haar distributed ensemble:
\begin{equation}
Expr = D_{KL}(\hat{P}_{PQC}(F; \Theta)\| P_{Haar}(F)).
\end{equation}
A smaller value for the KL divergence indicates a better ability to explore the Hilbert space. This measure of expressibility is observed on a logarithmic scale. This is where we decided to deviate from the original definition, as we will include these characteristics into the measure itself. In our work, we will evaluate the negative logarithmic of the KL divergence between the ensemble of Haar random states and the estimated fidelity distribution of the PQC, so that larger values for expressibility' correspond to better ability to explore the Hilbert space, and will be correlated on this logarithmic scale. We will refer to this as \emph{expressibility'} (\emph{Expr'}), distinguished by the ' symbol:
\begin{equation}
Expr' = - \log_{10}(D_{kl}(\hat{P}_{PQC}(F; \Theta)\| P_{Haar}(F))).
\end{equation}

In the same paper, \citet{th_sukin_2019} define a second descriptor named \emph{entangling capability}. This descriptor is intended to capture the entangling capability of a circuit which, based on prior art such as work by \citet{th_schuld_2018} and \citet{th_kandala_hevqe}, allow a PQC to capture non-trivial correlations in the quantum data. Multiple ways to compute this measure exist, but the Meyer-Wallach entanglement measure\citep{th_ge}, denoted Q, is chosen due to its scalability and ease of computation. It is defined as
\begin{equation}
    Q(\ket{\psi}) \equiv \frac{4}{n} \sum_{j=1}^{n} D(\iota_j(0) \ket{\psi}, \iota_j(1)\ket{\psi})
\end{equation}
Where $\iota_j(b)$ represents a linear mapping for a system of $n$ qubits that acts on the computational basis with $b_j \in \{0,1\}$:
\begin{equation}
    \iota_j(b)\ket{b_1 \dotsc b_n} = \delta_{bb_j} \ket{b_1 \dotsc \hat{b}_j \dotsc b_n}.
\end{equation}
Here, the symbol $\hat{ }$ denotes the absence of a qubit. In practice, also this measure of the PQC needs to be approximated by sampling, so \citet{th_sukin_2019} define the estimate of \emph{entangling capability} (\emph{Ent}) of the PQC to be the following:
\begin{equation}
    Ent = \frac{1}{|S|} \sum_{\theta_i \in S} Q(\ket{\psi_{\theta_i}}),
\end{equation}
where $S$ represents the set of sampled circuit parameter vectors $\theta$.

The quantification of both expressibility and entangling capability for the circuits, as found and provided by \citet{th_sukin_2019}, is shown in Table \ref{tab:expr_ent_tab} in Appendix \ref{sec:expr_ent}. 

\subsection{Datasets}
\begin{figure}[hbt]
	\centering
	\includegraphics[width=\columnwidth]{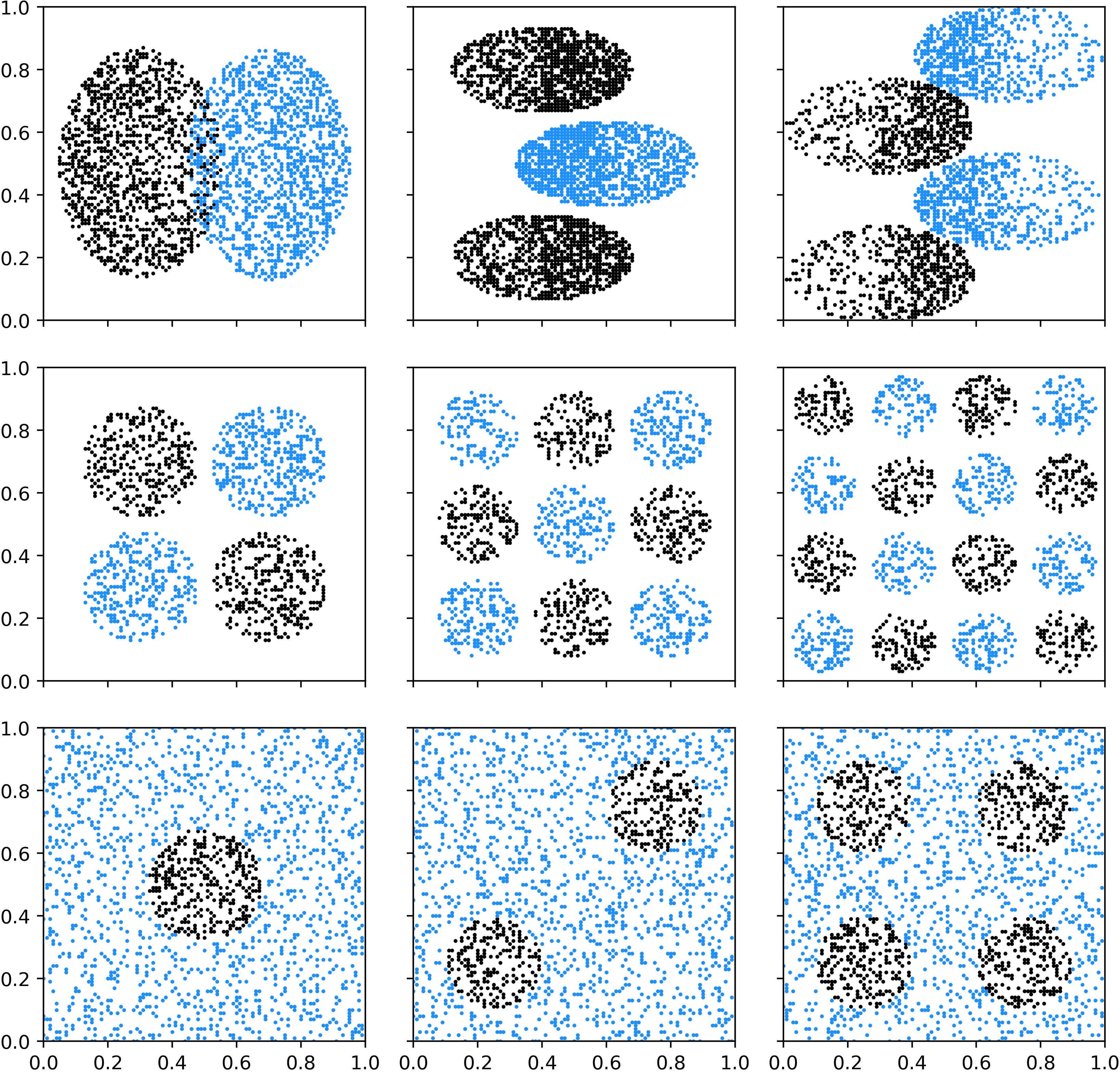}
    \caption{Datasets used to benchmark classification accuracy, labelled numerically in the vertical direction and alphabetically in the horizontal direction}
	\label{fig:dataset}  
	\vspace*{-1em}
\end{figure}

Many datasets for the evaluation of classical Machine Learning algorithms exist\citep{th_datasets}. However, classical Machine Learning is more advanced as a field compared to Quantum Machine Learning, and currently capable of both processing data at larger scales and predicting more complex distributions. We have searched for a set of problems of increasing and varying difficulty, but not of larger problem size, to cover a wider range of problems to benchmark against. We were unable to find any that satisfied our needs, and therefor came up with a set of nine datasets for the classifiers to fit, as shown in Figure \ref{fig:dataset}, labelled numerically in the vertical direction and alphabetically in the horizontal direction. Here, we included datasets where the two classes are bordering one another (1a), require one (1b) or more (1c) bends in a decision boundary, entrap one another (2a, 2b, 2c) or fully encapsulate each other (3a, 3b, 3c). Each dataset contains a total of $1500$ data points for training, testing hyperparameters, and validation in a ratio of $3:1:1$. The ratio of data points labelled true versus data points labelled false is $1:1$, e.g. all datasets are balanced. All data points are cleaned and normalized.

\subsection{Embedding the data}
\begin{figure}[b!]
	\centering
	\includegraphics[width=\columnwidth]{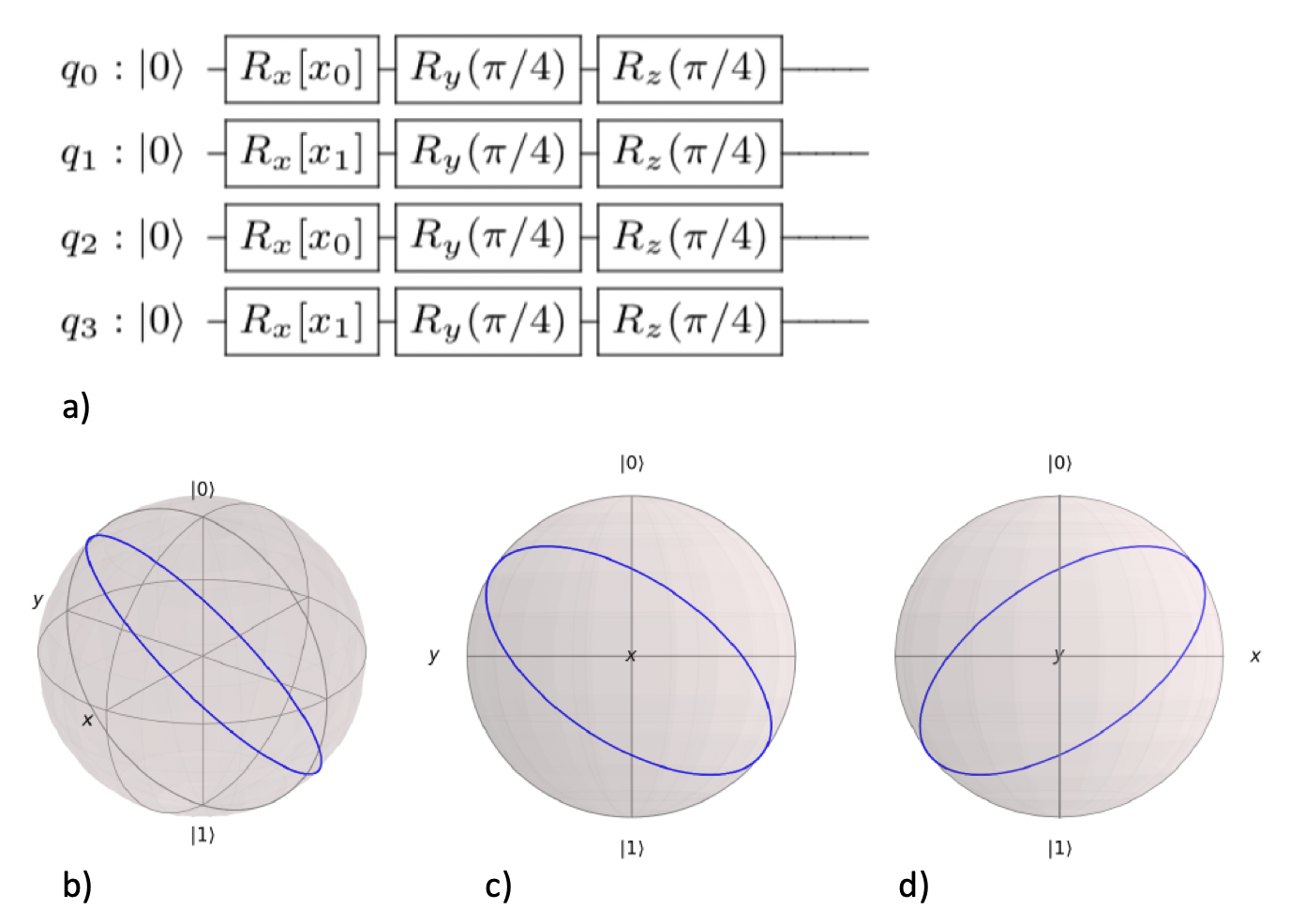}
    \caption{The various aspects of the minimally expressive embedding: a) the embedding circuits; b) the quantum state achieved when uniformly sampling input data, viewed in 3d; c,d): the quantum state achieved when uniformly sampling input data, in 2d on y-z and in 2d on x-z}
	\label{fig:min_expressive_embedding}  
	\vspace*{-1em}
\end{figure}

Various ways to embed classical data into quantum circuits have been proposed, such as Amplitude Encoding \citep{th_schuld_2019_qmlhs}, Product Encoding\citep{th_stoudenmire_2016}, or Squeezed Vacuum State Encoding\citep{th_schuld_2019_qmlhs}. Important distinguishing characteristics are time complexity, memory complexity and fit for prediction circuits, as some prediction algorithms require data in a particular format. This part of the QNN is also referred to as the \emph{state preparation circuit}, \emph{embedding circuit} or the \emph{feature embedding circuit}. However, the existing circuits do not meet our requirements. We search for an embedding that
\begin{itemize}
    \item holds as little expressive power as possible, as we want to observe the expressive power of the PQC
    \item does not create a bias in regard to the computational bases the PQC operates on
\end{itemize}
For this, we propose a novel embedding that we will refer to as the \emph{minimally expressive embedding}. To make sure this embedding circuit holds as little expressive power as possible, we embed the classical data into a quantum state using a linear mapping of a single parameterized Pauli X gate. This can be visualized as embedding data along a circle in the Hilbert space, as shown in Figure \ref{fig:min_expressive_embedding}b. We then use a Pauli Y and Z gate to rotate the circle $45$ degrees both in X and Y. The result is an embedding that, when projected down on the various computational basis, can address a similar range in every computational basis. This can be seen in Figure \ref{fig:min_expressive_embedding}c and \ref{fig:min_expressive_embedding}d.

We embed the 2-dimensional data in a replicative fashion: data on the x-axis in Figure \ref{fig:dataset} is embedded in qubit 0 and 2, data on the y-axis is embedded in qubit 1 and 3. We do this for two reasons:
\begin{itemize}
    \item 2-dimensional data can be easily visualized for a better understanding of both the data and the fitted decision boundary
    \item input redundancy is suggested to provide an advantage in classification accuracy\citep{th_input_redun}
\end{itemize}
The full embedding circuit is shown in Figure \ref{fig:min_expressive_embedding}a. 

\subsection{The Parameterized Quantum Circuit} 
In this paper, we aim to explore the correlation between our previously introduced descriptors and classification accuracy for various circuits and layers. For this, we will evaluate the same circuits that are used by \citet{th_sukin_2019}. These circuits consist of rotational gates, parameterized rotational gates and conditional rotational gates. Every experiment follows the same design: a single embedding circuit followed by 1 or 2 circuit templates. The exact layout of the circuit templates can be consulted in Figure \ref{fig:circuits} in Appendix \ref{sec:circuits}. All 19 circuits, except circuit 10, are used in the hyperparameter selection runs. All 19 circuits are used in the validation runs. 

\subsection{Measurement observable and mapping}
The circuit needs to be measured repeatedly. The number of times can be significantly less as previously thought\citep{th_stoch_gd}. We use the Pauli Z as an observable on all four qubits. This results in a total of 16 possible states, of which we map the first four and last four to output class $-1$ and the other to output class $1$. This balanced mapping aligns with the balance between the true and false data points in our dataset.

\subsection{Loss function} 
Loss functions quantify the error between the predicted class of the sample and the actual class. This can be done in several ways, which creates different loss landscapes. These loss landscapes have different properties for different loss functions, example properties being plateaus, local minima, and global minima that do not align with the true minima. In our work, we will evaluate the \emph{L1} loss and \emph{L2} loss. 

The L1 loss can be seen as the most straight-forward loss function, taking the absolute value of the difference between the $y_{true}$ and the $y_{pred}$. However, taking the absolute value is not preferred, as it is difficult to differentiate. The L1 loss is defined as follows:
\begin{equation}
    L_1 = \displaystyle\sum_{s=0}^{samples}|y_{pred_s} - y_{true_s}|.
\end{equation}

The L2 loss does not use the absolute value but instead ensures positive outcomes through taking the square of the difference, which is easier to differentiate. This square also penalizes large errors harder than small errors, reducing the chance of overfitting. The L2 loss is defined as follows: 
\begin{equation}
    L_2 = \displaystyle\sum_{s=0}^{samples}(y_{pred_s} - y_{true_s})^2.
\end{equation}
For binary values of y, the L2 loss and the L1 loss are equal. This is not the case for continuous values. 

\subsection{Optimizer}
The task of the optimizer is to find updates to the parameters based on the outcome of the loss function in such a way that, after repeated runs, the loss is minimized. Many approaches make use of the gradient, either analytically \citep{th_analytic_gd} or approximately\citep{th_stoch_gd}. Alternative approaches exist, such as genetic optimization strategies\citep{th_swarm_opt}. Either way, a balance needs to be struck between making large enough jumps to get out of local minima and plateaus, and making small enough updates so that the jumps don't equal random walks in the loss landscape. In our work, we evaluate the \emph{Adam} optimizer and the \emph{Gradient Descent} optimizer, due to their popularity in classical frameworks and their availability in our implementation framework. 

\subsection{Implementation framework}
We originally implemented our work in Qiskit\citep{th_qiskit}, but switched to Pennylane\citep{th_pennylane} due to the ease for our hyperparameter search.

\subsection{Training setup}
We will first perform a hyperparameter search on the loss functions L1 and L2, as well as on the optimizers Adam and Gradient Descent, all introduced previously in this section. We compare their performance in terms of average classification accuracy over the test data of our nine datasets. This will require a total of 1368 simulations. We will report on the difference in accuracy for different numbers of layers of each circuit but as each layer configuration has its own value for the descriptors, we will treat each layer configuration as unique data points in our final analysis. The optimal hyperparameters derived through the test runs are used as settings for the validation runs during which we gather our final accuracy values for all 19 circuits in both 1 and 2 layer configurations. As a sanity check, we will repeat the final experiment three times, requiring a total of 1026 additional simulations. 

\subsection{Defining correlation}
In order to determine the correlation, we will calculate the Pearson Product-Moment Correlation Coefficient\citep{th_pearson} between expressibility and classification accuracy, as well as between entangling capability and classification accuracy on the 342 data points, as described in the previous section. We will use these coefficients to draw conclusions on the level of correlation between the descriptors and the classification accuracy. 

\subsection{Classical Neural Network}    
We will also evaluate our dataset using a classical Neural Network for comparison and sanity checking. We implemented both a 1 and 2 layer version, each having 16 weights per layer. The activation function used is the \emph{Rectified Linear Unit} (\emph{ReLU}), and the system is optimized using the Adam optimizer. All is implemented in \emph{Tensorflow}\citep{th_tf}.

\section{\bf{Results}} \label{sec:results}

\begin{figure*}[h!]
     \centering
     \begin{subfigure}[b]{0.32\textwidth}
         \centering
         \includegraphics[width=\textwidth]{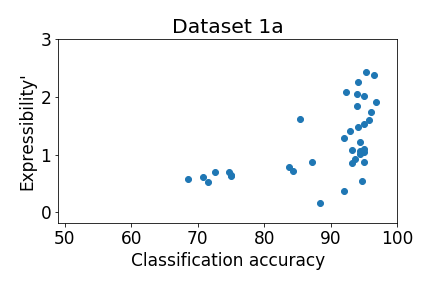}
     \end{subfigure}
     \hfill
     \begin{subfigure}[b]{0.32\textwidth}
         \centering
         \includegraphics[width=\textwidth]{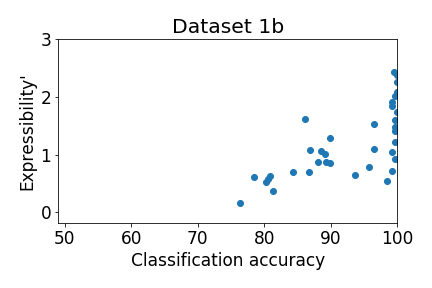}
     \end{subfigure}
     \hfill
     \begin{subfigure}[b]{0.32\textwidth}
         \centering
         \includegraphics[width=\textwidth]{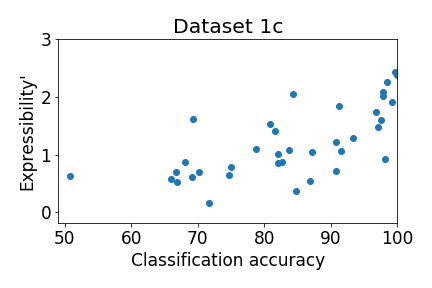}
     \end{subfigure}
     
     \begin{subfigure}[b]{0.32\textwidth}
         \centering
         \includegraphics[width=\textwidth]{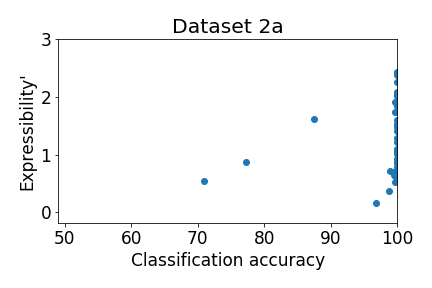}
     \end{subfigure}
     \hfill
     \begin{subfigure}[b]{0.32\textwidth}
         \centering
         \includegraphics[width=\textwidth]{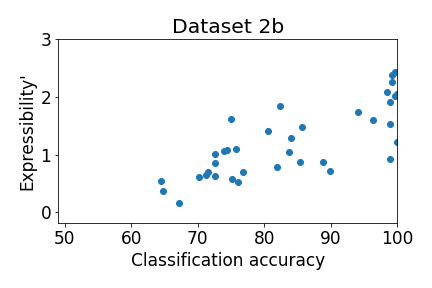}
     \end{subfigure}
     \hfill
     \begin{subfigure}[b]{0.32\textwidth}
         \centering
         \includegraphics[width=\textwidth]{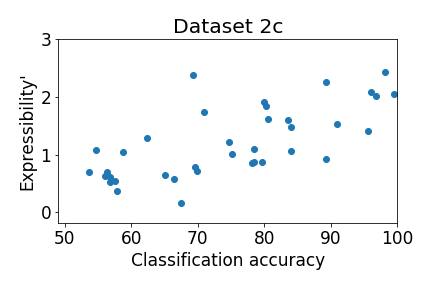}
     \end{subfigure}
     
     \begin{subfigure}[b]{0.325\textwidth}
         \centering
         \includegraphics[width=\textwidth]{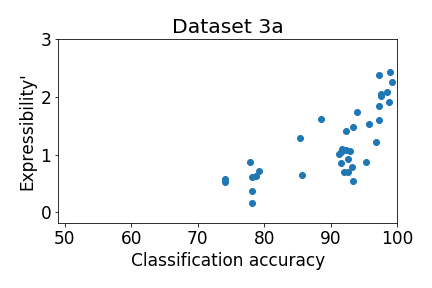}
     \end{subfigure}
     \hfill
     \begin{subfigure}[b]{0.325\textwidth}
         \centering
         \includegraphics[width=\textwidth]{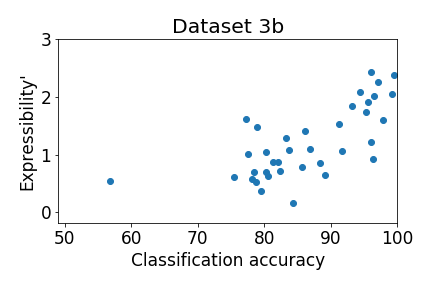}
     \end{subfigure}
     \hfill
     \begin{subfigure}[b]{0.325\textwidth}
         \centering
         \includegraphics[width=\textwidth]{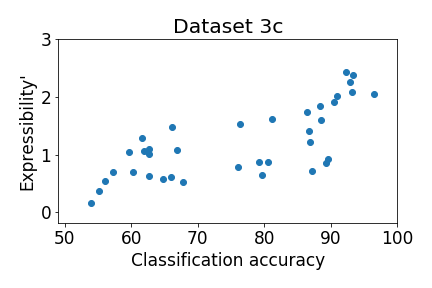}
     \end{subfigure}
    \caption{Classification accuracy versus expressibility'}
    \label{fig:corr_expr}
\end{figure*}

\begin{figure*}[h!]
     \centering
     \begin{subfigure}[b]{0.32\textwidth}
         \centering
         \includegraphics[width=\textwidth]{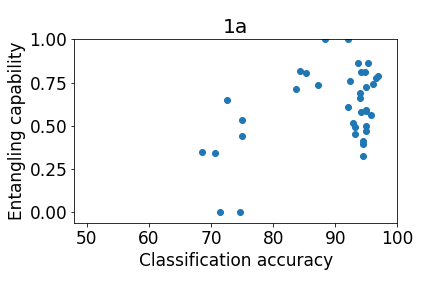}
     \end{subfigure}
     \hfill
     \begin{subfigure}[b]{0.32\textwidth}
         \centering
         \includegraphics[width=\textwidth]{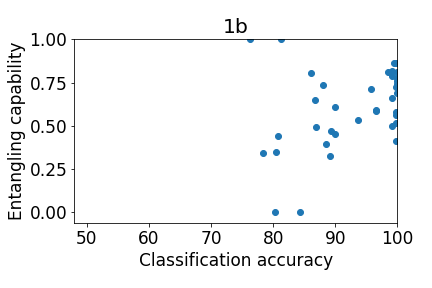}
     \end{subfigure}
     \hfill
     \begin{subfigure}[b]{0.32\textwidth}
         \centering
         \includegraphics[width=\textwidth]{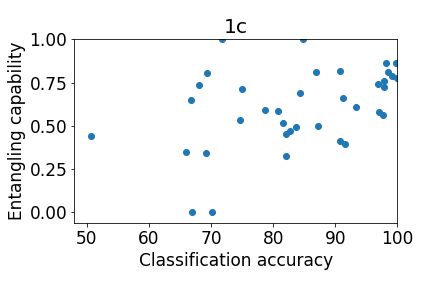}
     \end{subfigure}
     
     \begin{subfigure}[b]{0.32\textwidth}
         \centering
         \includegraphics[width=\textwidth]{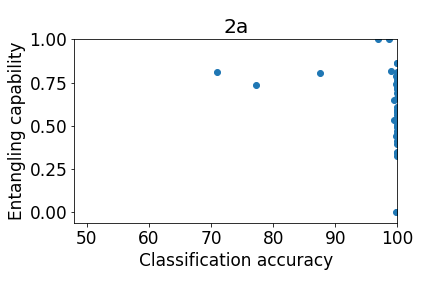}
     \end{subfigure}
     \hfill
     \begin{subfigure}[b]{0.32\textwidth}
         \centering
         \includegraphics[width=\textwidth]{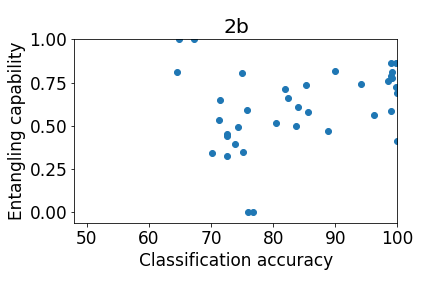}
     \end{subfigure}
     \hfill
     \begin{subfigure}[b]{0.32\textwidth}
         \centering
         \includegraphics[width=\textwidth]{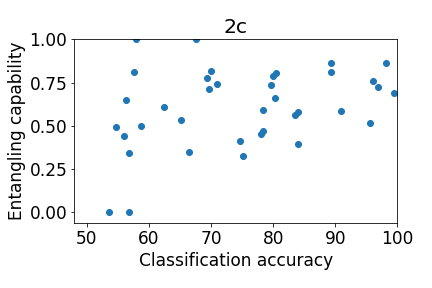}
     \end{subfigure}
     
     \begin{subfigure}[b]{0.325\textwidth}
         \centering
         \includegraphics[width=\textwidth]{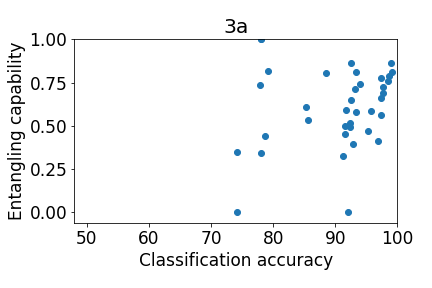}
     \end{subfigure}
     \hfill
     \begin{subfigure}[b]{0.325\textwidth}
         \centering
         \includegraphics[width=\textwidth]{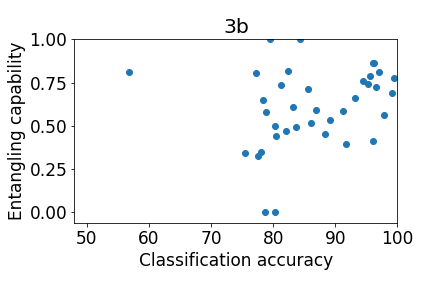}
     \end{subfigure}
     \hfill
     \begin{subfigure}[b]{0.325\textwidth}
         \centering
         \includegraphics[width=\textwidth]{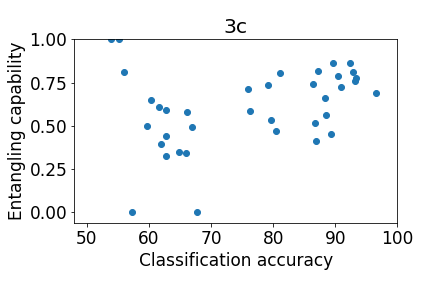}
     \end{subfigure}
    \caption{Classification accuracy versus entangling capability}
	\label{fig:corr_ent}  
\end{figure*}

\begin{table}[b]
\caption{Accuracy for various hyperparameter settings, averaged out over the datasets.}
\label{tab:hyperparameters}
    \begin{tabularx}{8cm}{X|X|X|X}
    \diagbox[width=2cm]{\textbf{Opt.}}{\textbf{Layers}} &\centering 1 &\centering 2 &      \diagbox[dir=SW,width=2cm]{\textbf{Layers}}{\textbf{Loss}} \\ \hline
    \multirow{2}{\hsize}{Adam} &\centering $72.6$ &\centering $76.7$ & $L1$ \\ \cline{2-4}
     &\centering $76.8$ &\centering $\boldsymbol{82.6}$ & L2 \\ \hline
    \multirow{2}{\hsize}{Gradient\\ decent} &\centering $73.9$ &\centering $71.5$ & $L1$ \\ \cline{2-4}
     &\centering $74.2$ &\centering $75.6$ & $L2$ \\ \hline
    \end{tabularx}
\end{table}
The classification accuracy for the various hyperparameter settings can be found in Appendix \ref{sec:app_hyp}. In particular, the results for the Adam optimizer with L1 loss can be found in Table \ref{tab:adam_l1} and with L2 loss in Table \ref{tab:adam_l2}. The results for the Gradient Descent optimizer with L1 loss in Table \ref{tab:gd_l1} and with L2 loss in Table \ref{tab:gd_l2}. The average classification accuracy across all datasets for the various hyperparameter settings, as well as the number of layers, are summarized in Table \ref{tab:hyperparameters}. Here, we see that the Adam optimizer combined with L2 loss achieves the best classification accuracy, regardless of the number of layers. Treating each hyperparameter separately using the \emph{factorial design}\citep{th_fact_design}, as shown in table \ref{tab:dep_var}, reconfirmed these settings. The outcome of the three validation runs using the L2 loss and Adam optimizer can be found in Appendix \ref{sec:app_val}, Table \ref{tab:val_1l} for the 1 layer runs and Table \ref{tab:val_2l} for the 2 layer runs. We calculated the mean absolute difference between each of the 342 data points of every run, and found it to be $0.06$ and a standard deviation of $0.36$. With this info, we expect the correlation not to vary significantly between the runs, which is also what we observe in Table \ref{tab:corr}, where we show the Pearson Product-Moment Correlation Coefficients between the descriptors and classification accuracy for each dataset individually, as well as the mean and standard deviation. We added an extra row where we exclude dataset 2a for reasons described in the discussion section. The relation between expressibility' and classification accuracy for every dataset is plotted in Figure \ref{fig:corr_expr}, the relation between entangling capability and classification accuracy is plotted in Figure \ref{fig:corr_ent}. The average classification accuracy for the classical NN is shown in Table \ref{tab:cnn}. 

\begin{table}[t]
\caption{Factorial design to evaluate classification accuracy ("Acc") with regard to the dependent variables ("DV")}
\label{tab:dep_var}
    \begin{tabularx}{8cm}{XX|XX|XX}
    \hline
    \textbf{DV} && \textbf{Option} & \textbf{Acc} & \textbf{Option} & \textbf{Acc}  \\
    \hline
    Loss && L1 & $73.7$ & $L2$ & $\boldsymbol{77.3}$ \\
    Optimizer && Adam & $\boldsymbol{77.2}$ & $GD$ & $73.8$ \\
    Layers && 1 & $74.4$ & $2$ & $\boldsymbol{76.7}$\\
    \hline
\end{tabularx}
\end{table}

\begin{table}[b]
\caption{Pearson Product-Moment Correlation Coefficient between Expressibility' and Classification Accuracy, as well as between Entangling Capability and Classification Accuracy, for the various datasets. Mean and standard deviation are taken over all datasets, mean' and standard deviation' is taken over all datasets but dataset 2a.}
\label{tab:corr}
\begin{tabularx}{8.8cm}{X|XXX|XXX}
\hline
 & \multicolumn{3}{c|}{\textbf{Expr' vs Acc}}  & \multicolumn{3}{c}{\textbf{Ent vs. Acc}}\\
 
Dataset & Run $1$ & Run $2$ & Run $3$ & Run $1$ & Run $2$ & Run $3$ \\
\hline
1a & 0.575 & 0.570 & 0.571 & 0.467 & 0.463 & 0.466 \\
1b & 0.699 & 0.699 & 0.699 & 0.353 & 0.353 & 0.353 \\
1c & 0.675 & 0.678 & 0.676 & 0.419 & 0.421 & 0.420 \\
2a & 0.200 & 0.200 & 0.200 & -0.257 & -0.257 & -0.258 \\
2b & 0.761 & 0.761 & 0.777 & 0.251 & 0.251 & 0.258 \\
2c & 0.700 & 0.707 & 0.694 & 0.339 & 0.343 & 0.336 \\
3a & 0.732 & 0.731 & 0.740 & 0.190 & 0.189 & 0.195 \\
3b & 0.693 & 0.686 & 0.693 & 0.231 & 0.226 & 0.231 \\
3c & 0.727 & 0.730 & 0.730 & 0.301 & 0.303 & 0.303 \\ \hline
Mean & 0.640 & 0.640 & 0.642 & 0.255 & 0.255 & 0.256 \\
Stdev & 0.163 & 0.164 & 0.165 & 0.199 & 0.199 & 0.199 \\ \hline
Mean' & 0.695 & 0.695 & 0.697 & 0.319 & 0.319 & 0.320 \\
Stdev' & 0.052 & 0.054 & 0.057 & 0.089 & 0.089 & 0.087 \\
\hline
\end{tabularx}
\end{table}

\section{\bf{Discussion}} \label{sec:discussion}

\subsection{Limitations of the experiment}
Before discussing the results of our experiment, we will address the limiting factors to place the finding in the correct perspective. 
\begin{itemize}
    \item Our experiment only includes specific quantum circuit designs. In particular, the system always starts with an embedding circuit, followed by 1 or 2 circuit templates, with rotational gates, parameterized rotational gates, and conditional rotational gates
    \item We handcrafted 9 different sets of data points of increasing difficulty that we believe to be realistic and representative problems for current-day Quantum Machine Learning algorithms. However, the sets only contain classical data with 2 features. Data encapsulating more complex patterns, or higher-dimensional data, might yield different results
    \item The 2d data was embedded in a replicative manner on 4 qubits with our proposed minimally expressive embedding circuit. Different embeddings, additional embeddings, or repetitive embeddings are not evaluated. Neither were ancilla qubits.
\end{itemize}
Furthermore, different descriptors for expressibility and entangling capability exist, as well as different descriptors for the power of a PQC overall\citep{th_brian, th_sukin_2019, th_schuld_2020_de}.

\subsection{Correlation}
For our experimental setup, we have found an average Pearson Correlation Coefficient between expressibility' and classification accuracy of $0.64\pm0.16$, as calculated from run 1 in Table \ref{tab:corr}. However, when looking at the plots in Figure \ref{fig:corr_expr}, we see that dataset $2a$ contains 33 out of 38 data points close to or at $100\%$ accuracy. As these accuracies are capped out, no meaningful correlation can be expected. For this reason, we mark dataset $2a$ as faulty and exclude it from our evaluation. This brings us to a final mean Pearson Correlation Coefficient between expressibility' and classification accuracy of $-0.70\pm0.05$. This indicates a strong correlation\citep{th_pcc_interpreatation} between classification accuracy and expressibility'. 

Using the same experimental setup, the mean Pearson Correlation Coefficient between entangling capability and classification accuracy across all datasets is $0.26\pm0.20$. As this experiment suffers the same characteristics for dataset $2a$, as observed in Figure \ref{fig:corr_ent}, we also decide to exclude dataset $2a$ for our evaluation of this coefficient. This brings the final mean Pearson Correlation Coefficient between entangling capability and classification accuracy at $0.32\pm0.09$. This indicates a weak correlation\citep{th_pcc_interpreatation} between classification accuracy and entangling capability.  

\begin{table}[b!]
\caption{Mean Pearson Correlation Coefficient for expressibility and classification accuracy as a sanity check on the remaining hyperparameters}
\label{tab:p_hyp}
\begin{tabularx}{8cm}{X|X|X|X}
\hline
\textbf{Optimizer} & \textbf{Loss} & \textbf{Mean} & \textbf{Stdev} \\ \hline
Adam & L1 & 0.63 & 0.13 \\ 
Adam & L2 & 0.70 & 0.08 \\
GD & L1 & 0.46 & 0.11 \\
GD & L2 & 0.56 & 0.13 \\ \hline
\end{tabularx}
\end{table}

\subsection{Experimental setup}
As another sanity check, we calculated the mean Pearson Correlation Coefficient and its standard deviation for the different hyperparameter settings on the test data, all with dataset $2a$ excluded. This is summarized in Table \ref{tab:p_hyp}. Here we see that for the Adam optimizer with either L1 or L2, we maintain a value that can be classified as a strong correlation. For Gradient Descent, the claim would be weaker, being classified as a moderate to strong correlation\citep{th_pcc_interpreatation}. After examining the data, we see that this is caused by the optimizer not performing optimally. In particular, at least half of the circuits on dataset $2c$ and $3c$ cannot fit any data, staying around $50\%$ accuracy, and the majority of the circuits are stuck in a local minima around $80\%$ for dataset $3b$. This is not surprising, as we saw in Table \ref{tab:dep_var} that the average classification accuracy while using the Gradient Descent optimizer is lower compared to using the Adam optimizer.

\subsection{Limitations of the descriptor} 
The original paper by \citet{th_sukin_2019} contains 19 circuits with 1  to 5 layers. In their paper, they address that expressibility values appear to saturate for all circuits, albeit at different levels. The lowest value of expressibility that they present is $0.0026$, which in our measure of expressibility' corresponds to $-log_{10}(0.0026) = 2.59$. In our experiments, we tested 6 circuits that have a value of expressibility' higher than 2, meaning an expressibility smaller than $0.01$. The best average accuracy achieved by any of our circuits is $97.7$. Still, this includes a fitting of only $92.3\%$ for dataset $3b$. In comparison, a classical Neural Network containing 2 layers with 16 weights is able to achieve an average classification accuracy of $99\%$ without any hyperparameter tuning, as shown in Table \ref{tab:cnn}. When looking at state-of-the-art Neural Network architectures, we continue to see these patterns. Most quantum classifiers are still being evaluated on small datasets such as ours\citep{th_schuld_2019_qmlhs, th_havlicek_2018}, or datasets such as the MNIST\citep{th_mnist}. Classical machine learning models on the other hand are being evaluated on larger and more complex datasets, such as ImageNet\citep{th_imagenet}. Although this may be purely due to the infancy of the current systems, limiting the size of the input data, it still appears that adding extra layers does not circumvent the saturation. We believe a hint might lay in the following reasoning: it is not only important for a classifier to be able to uniformly address a large Hilbert space, but also requires a repeated non-linear mapping between these spaces. As a thought experiment, think of a circuit that embeds linearly increasing classical data with a single rotational Pauli X gate. In this scenario, one would not expect to be able to find a separating plane between the two classes without remapping the data. We believe that this is also the reason why in classical Neural Networks, the data is repeatedly mapped between feature spaces by repeated layers of linear neurons and non-linear activation functions. We believe that quantum circuits need to be designed and evaluated in such a manner too. Recent research addresses this by having alternating layers of embedding and trainable circuit layers\citep{th_alternation}, thereby breaking linearity, although \citet{th_schuld_2020_de} they drive a point that classical systems can achieve a similar effect with similar resources, deeming the use of quantum for ML not necessary. Alternatively, one could envision the use of intermediary measurements at the end of a circuit template to collapses the quantum state, followed by the next circuit template, creating similar effects to those of an activation function. The investigation into the design and use as a measure for quantification is marked as future work. 

\subsection{Threats to validity}
We are aware that different hyperparameters, different initialization of the weights and different datasets can yield different results. We have tried to account for this effect by performing a hyperparameter search on test data, and using the best hyperparameter settings to create the data points on the validation set. Still, as a sanity check, we analyzed the outcome on the test data containing the other hyperparameter settings too. Furthermore, we repeated our experiments three times to account for different initializations of the weights, and designed 9 different datasets of increasing difficulty. 

\section{\bf{Conclusion}} \label{sec:conclusion}
In our work, we have found a strong correlation between classification accuracy and expressibility', based on a mean $0.7\pm0.05$ Pearson Product-Moment Correlation Coefficient\citep{th_pcc_interpreatation}. This numerically derived outcome is calculated using 342 data points. These data points were generated using 19 circuits, in both 1 and 2 layer configurations, evaluated on 9 custom datasets of increasing difficulty. Here, expressibility' is based on the definition of expressibility proposed by \citet{th_sukin_2019}, and meant to capture the ability of a Parameterized Quantum Circuit in a hybrid quantum-classical framework to explore the Hilbert space. This is calculated by taking the negative log of the Kullback–Leibler divergence between the ensemble of Haar random states and the estimated fidelity distribution of the PQC. Our experiment is limited to PQCs that follow a specific pattern of concatenating an embedding layer with one or more circuit templates. It is suggested that this circuit setup is limiting the classification performance\citep{th_schuld_2020_de}. Further investigation into more elaborate designs that break linearity after embedding is required, for example by repeated alteration between embedding layers and trainable layers\citep{th_schuld_2020_de}. Such designs would potentially not be captured by the expressibility' measure, and further investigation into extending the measure is required. 

We have also investigated the correlation between entangling capability of a circuit and its classification accuracy, where entangling capability was measured as the Meyer-Wallach entanglement measure\citep{th_ge}. The outcome was a weak correlation, based on a similar experimental setup that yielded a mean Pearson Product-Moment Correlation of $0.32\pm0.09$\citep{th_pcc_interpreatation}.
\section{\bf{Acknowledgements}} \label{sec:acknowledgements}

The authors would like to thank Sukin Sim for providing the data points in Table \ref{tab:expr_ent_tab} and valuable feedback; Nicola Pancotti and Christoph Segler for providing feedback on our statistical approach; Jonas Haferkamp and Jordi Brugués for discussions on measures of expressibility and t-designs; and Ryan Sweke for discussions on function classes.


%
%

\bibliographystyle{spbasic}      

\bibliography{refs}
\balance

%
%

\clearpage
\onecolumn
\begin{appendices}

\section{\bf{Expressibility and entangling capability}} \label{sec:expr_ent}
\begin{table*}[ht]
    \caption{Expressibility, expressibility' and entangling\\ capability of the circuits. Please see copyright notice $^1$}
	\label{tab:expr_ent_tab}
	\begin{tabularx}{0.455\textwidth}{@{}l|lll|lll@{}}
	\hline
	 & \multicolumn{3}{c|}{\textbf{Layer 1}} & \multicolumn{3}{c}{\textbf{Layer 2}}\\
	Circuit & Expr & Expr' & Ent & Expr & Expr' & Ent \\
	\hline 
    $1$  & $ 0.2995 $ & $ 0.52 $ & $ 0.0 $ & $ 0.1972 $ & $ 0.71 $ & $ 0.0 $ \\
    $2$  & $ 0.2875 $ & $ 0.54 $ & $ 0.81 $ & $ 0.0244 $ & $ 1.61 $ & $ 0.8 $ \\
    $3$  & $ 0.24 $ & $ 0.62 $ & $ 0.34 $ & $ 0.0847 $ & $ 1.07 $ & $ 0.49 $ \\
    $4$  & $ 0.1353 $ & $ 0.87 $ & $ 0.47 $ & $ 0.0291 $ & $ 1.54 $ & $ 0.59 $ \\
    $5$  & $ 0.0601 $ & $ 1.22 $ & $ 0.41 $ & $ 0.0087 $ & $ 2.06 $ & $ 0.69 $ \\
    $6$  & $ 0.0041 $ & $ 2.38 $ & $ 0.78 $ & $ 0.0036 $ & $ 2.44 $ & $ 0.86 $ \\
    $7$  & $ 0.0985 $ & $ 1.01 $ & $ 0.33 $ & $ 0.0386 $ & $ 1.41 $ & $ 0.52 $ \\
    $8$  & $ 0.0864 $ & $ 1.06 $ & $ 0.39 $ & $ 0.0255 $ & $ 1.59 $ & $ 0.56 $ \\
    $9$  & $ 0.678 $ & $ 0.17 $ & $ 1.0 $ & $ 0.4261 $ & $ 0.37 $ & $ 1.0 $ \\
    $10$  & $ 0.2284 $ & $ 0.64 $ & $ 0.54 $ & $ 0.1617 $ & $ 0.79 $ & $ 0.71 $ \\
    $11$  & $ 0.1325 $ & $ 0.88 $ & $ 0.73 $ & $ 0.0122 $ & $ 1.92 $ & $ 0.79 $ \\
    $12$  & $ 0.2003 $ & $ 0.7 $ & $ 0.65 $ & $ 0.0181 $ & $ 1.74 $ & $ 0.74 $ \\
    $13$  & $ 0.0516 $ & $ 1.29 $ & $ 0.61 $ & $ 0.0083 $ & $ 2.08 $ & $ 0.76 $ \\
    $14$  & $ 0.0144 $ & $ 1.84 $ & $ 0.66 $ & $ 0.0055 $ & $ 2.26 $ & $ 0.81 $ \\
    $15$  & $ 0.191 $ & $ 0.72 $ & $ 0.82 $ & $ 0.1185 $ & $ 0.93 $ & $ 0.86 $ \\
    $16$  & $ 0.2615 $ & $ 0.58 $ & $ 0.35 $ & $ 0.0885 $ & $ 1.05 $ & $ 0.5 $ \\
    $17$  & $ 0.1378 $ & $ 0.86 $ & $ 0.45 $ & $ 0.0327 $ & $ 1.49 $ & $ 0.58 $ \\
    $18$  & $ 0.2358 $ & $ 0.63 $ & $ 0.44 $ & $ 0.0602 $ & $ 1.22 $ & $ 0.62 $ \\
    $19$  & $ 0.0814 $ & $ 1.09 $ & $ 0.59 $ & $ 0.0096 $ & $ 2.02 $ & $ 0.72 $ \\
    \hline
    \end{tabularx}
\end{table*}
$^1$Notice regarding data in the "Expr" and "Ent" columns: \\
Copyright Wiley-VCH GmbH. Reproduced with permission.\\
Source: Sukin Sim, Peter D. Johnson, and Alán Aspuru‐Guzik. "Expressibility and Entangling Capability of Parameterized Quantum Circuits for Hybrid Quantum‐Classical Algorithms." Advanced Quantum Technologies 2.12 (2019): 1900070. Page 9. 
    
\clearpage
\section{Circuits} \label{sec:circuits}
\begin{figure}[hbt]
	\centering
	\includegraphics[width=0.81\columnwidth]{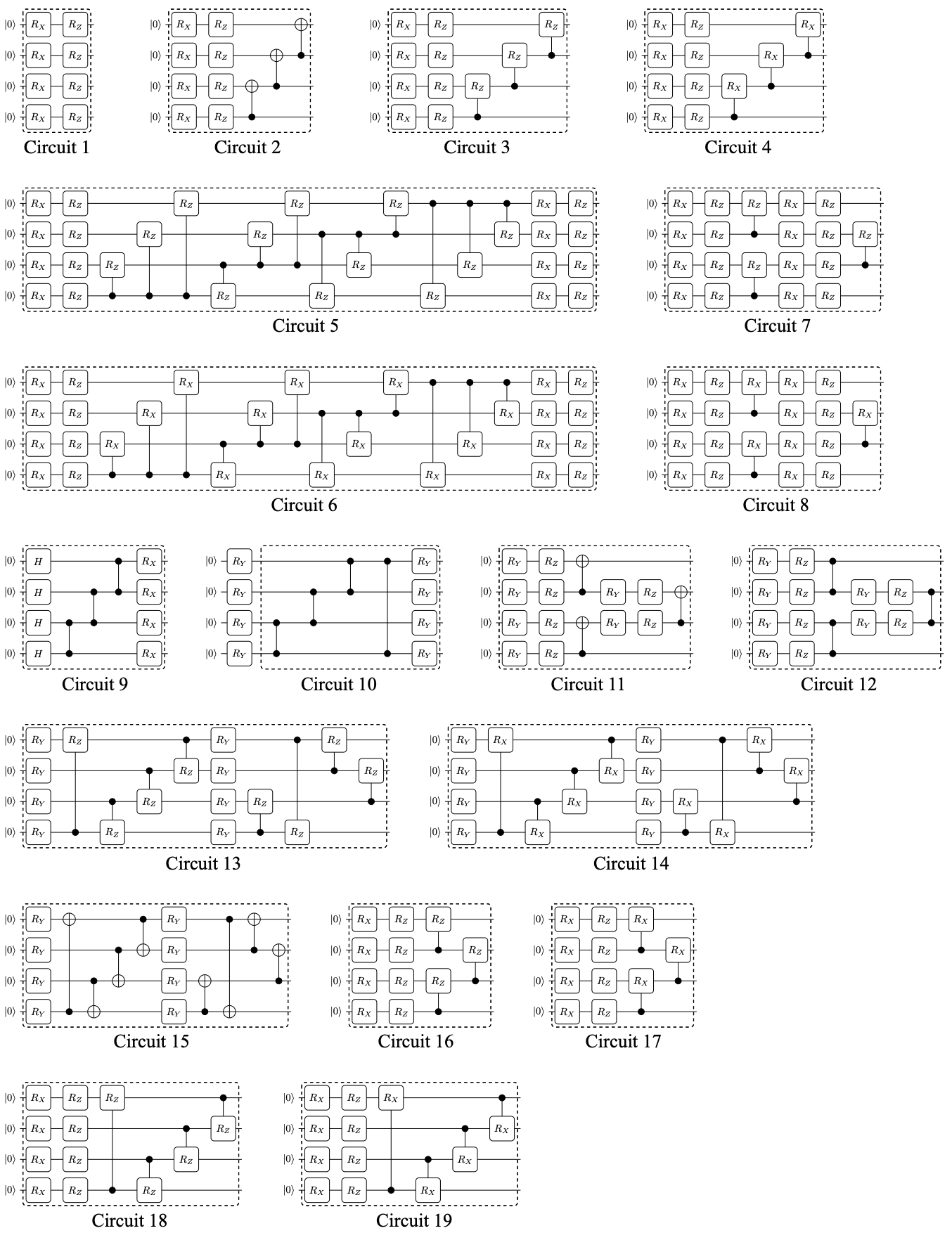}
    \caption{Circuit templates that are evaluated in this study. Circuit diagrams were generated using qpic\citep{th_qpic}. Please see copyright notice$^2$}
	\label{fig:circuits}  
	\vspace*{-1em}
\end{figure}
$^2$Notice regarding choice of circuit templates and visualizations of these circuits:\\
Copyright Wiley-VCH GmbH. Reproduced with permission.\\
Source: Sukin Sim, Peter D. Johnson, and Alán Aspuru‐Guzik. "Expressibility and Entangling Capability of Parameterized Quantum Circuits for Hybrid Quantum‐Classical Algorithms." Advanced Quantum Technologies 2.12 (2019): 1900070. Page 8. 

\clearpage
\section{\bf{Validation results}} \label{sec:app_val}
\begin{table*}[h!]
	\caption{Evaluation results --- classical Neural Network}
	\label{tab:cnn}
	\begin{tabularx}{0.66\textwidth}{@{}l|lllllllll|l@{}}
    \hline
	\textbf{Setup} & \multicolumn{10}{c}{\textbf{Performance}} \\
	\hline
    Layers & 1a & 1b & 1c & 2a & 2b & 2c & 3a & 3b & 3c & Avg. \\
	\hline 
    1 & $96\%$ & $100\%$ & $100\%$ & $92\%$ & $94\%$ & $82\%$ & $100\%$ & $98\%$ & $70\%$ & $92\%$ \\
	2 & $96\%$ & $100\%$ & $100\%$ & $100\%$ & $99\%$ & $98\%$ & $100\%$ & $99\%$ & $97\%$ & $99\%$ \\
	    \hline
    \end{tabularx}
\end{table*}

\begin{table*}[h!]
	\caption{Validation results --- Adam optimzer, L2 loss, 1 layer}
	\label{tab:val_1l}
	\begin{tabularx}{0.84\textwidth}{@{}ll|lllllllll|l@{}}
    \hline
	\multicolumn{2}{c|}{\textbf{Setup}} & 
	\multicolumn{10}{c}{\textbf{Performance}} \\
	\hline
	Run & Circuit & 1a & 1b & 1c & 2a & 2b & 2c & 3a & 3b & 3c & Avg. \\
	\hline 
	
$1$ & $1$ & $71.5\%$ & $80.3\%$ & $66.9\%$ & $99.7\%$ & $76.0\%$ & $56.8\%$ & $74.1\%$ & $78.7\%$ & $67.7\%$ & $74.6\%$\\
$1$ & $2$ & $94.7\%$ & $98.4\%$ & $86.9\%$ & $70.9\%$ & $64.5\%$ & $57.6\%$ & $93.3\%$ & $56.8\%$ & $56.0\%$ & $75.5\%$\\
$1$ & $3$ & $70.7\%$ & $78.4\%$ & $69.1\%$ & $100.0\%$ & $70.1\%$ & $56.8\%$ & $78.1\%$ & $75.5\%$ & $65.9\%$ & $73.8\%$\\
$1$ & $4$ & $94.9\%$ & $89.3\%$ & $82.7\%$ & $100.0\%$ & $88.8\%$ & $78.4\%$ & $95.2\%$ & $82.1\%$ & $80.5\%$ & $88.0\%$\\
$1$ & $5$ & $94.4\%$ & $99.7\%$ & $90.7\%$ & $100.0\%$ & $100.0\%$ & $74.7\%$ & $96.8\%$ & $96.0\%$ & $86.9\%$ & $93.2\%$\\
$1$ & $6$ & $96.5\%$ & $100.0\%$ & $100.0\%$ & $100.0\%$ & $99.2\%$ & $69.3\%$ & $97.3\%$ & $99.5\%$ & $93.3\%$ & $95.0\%$\\
$1$ & $7$ & $94.4\%$ & $89.1\%$ & $82.1\%$ & $100.0\%$ & $72.5\%$ & $75.2\%$ & $91.2\%$ & $77.6\%$ & $62.7\%$ & $82.8\%$\\
$1$ & $8$ & $94.4\%$ & $88.5\%$ & $91.5\%$ & $100.0\%$ & $73.9\%$ & $84.0\%$ & $92.8\%$ & $91.7\%$ & $61.9\%$ & $86.5\%$\\
$1$ & $9$ & $88.3\%$ & $76.3\%$ & $71.7\%$ & $96.8\%$ & $67.2\%$ & $67.5\%$ & $78.1\%$ & $84.3\%$ & $53.9\%$ & $76.0\%$\\
$1$ & $10$ & $74.9\%$ & $93.6\%$ & $74.7\%$ & $99.5\%$ & $71.2\%$ & $65.1\%$ & $85.6\%$ & $89.1\%$ & $79.7\%$ & $81.5\%$\\
$1$ & $11$ & $87.2\%$ & $88.0\%$ & $68.0\%$ & $77.3\%$ & $85.3\%$ & $79.7\%$ & $77.9\%$ & $81.3\%$ & $79.2\%$ & $80.4\%$\\
$1$ & $12$ & $72.5\%$ & $86.7\%$ & $66.7\%$ & $99.5\%$ & $71.5\%$ & $56.3\%$ & $92.5\%$ & $78.4\%$ & $60.3\%$ & $76.0\%$\\
$1$ & $13$ & $92.0\%$ & $89.9\%$ & $93.3\%$ & $100.0\%$ & $84.0\%$ & $62.4\%$ & $85.3\%$ & $83.2\%$ & $61.6\%$ & $83.5\%$\\
$1$ & $14$ & $93.9\%$ & $99.2\%$ & $91.2\%$ & $100.0\%$ & $82.4\%$ & $80.3\%$ & $97.3\%$ & $93.1\%$ & $88.3\%$ & $91.7\%$\\
$1$ & $15$ & $84.3\%$ & $99.2\%$ & $90.7\%$ & $98.9\%$ & $89.9\%$ & $69.9\%$ & $79.2\%$ & $82.4\%$ & $87.2\%$ & $86.8\%$\\
$1$ & $16$ & $68.5\%$ & $80.5\%$ & $65.9\%$ & $100.0\%$ & $75.2\%$ & $66.4\%$ & $74.1\%$ & $78.1\%$ & $64.8\%$ & $74.8\%$\\
$1$ & $17$ & $93.1\%$ & $89.9\%$ & $82.1\%$ & $100.0\%$ & $72.5\%$ & $78.1\%$ & $91.5\%$ & $88.3\%$ & $89.3\%$ & $87.2\%$\\
$1$ & $18$ & $74.9\%$ & $80.8\%$ & $50.7\%$ & $99.7\%$ & $72.5\%$ & $56.0\%$ & $78.7\%$ & $80.5\%$ & $62.7\%$ & $72.9\%$\\
$1$ & $19$ & $94.9\%$ & $96.5\%$ & $78.7\%$ & $100.0\%$ & $75.7\%$ & $78.4\%$ & $91.7\%$ & $86.9\%$ & $62.7\%$ & $85.1\%$\\ \hline

$2$ & $1$ & $71.5\%$ & $80.3\%$ & $66.9\%$ & $99.7\%$ & $76.0\%$ & $56.8\%$ & $74.1\%$ & $78.7\%$ & $67.7\%$ & $74.6\%$\\
$2$ & $2$ & $94.7\%$ & $98.4\%$ & $86.9\%$ & $70.9\%$ & $64.5\%$ & $57.6\%$ & $93.3\%$ & $56.8\%$ & $56.0\%$ & $75.5\%$\\
$2$ & $3$ & $70.7\%$ & $78.4\%$ & $69.1\%$ & $100.0\%$ & $70.1\%$ & $56.8\%$ & $78.1\%$ & $75.5\%$ & $65.9\%$ & $73.8\%$\\
$2$ & $4$ & $94.9\%$ & $89.3\%$ & $82.7\%$ & $100.0\%$ & $88.8\%$ & $78.4\%$ & $95.2\%$ & $82.1\%$ & $80.5\%$ & $88.0\%$\\
$2$ & $5$ & $94.4\%$ & $99.7\%$ & $90.7\%$ & $100.0\%$ & $100.0\%$ & $74.7\%$ & $96.8\%$ & $96.0\%$ & $86.9\%$ & $93.2\%$\\
$2$ & $6$ & $96.5\%$ & $100.0\%$ & $100.0\%$ & $100.0\%$ & $99.2\%$ & $69.3\%$ & $97.3\%$ & $99.5\%$ & $93.3\%$ & $95.0\%$\\
$2$ & $7$ & $94.4\%$ & $89.1\%$ & $82.1\%$ & $100.0\%$ & $72.5\%$ & $75.2\%$ & $91.2\%$ & $77.6\%$ & $62.7\%$ & $82.8\%$\\
$2$ & $8$ & $94.4\%$ & $88.5\%$ & $91.5\%$ & $100.0\%$ & $73.9\%$ & $84.0\%$ & $92.8\%$ & $91.7\%$ & $61.9\%$ & $86.5\%$\\
$2$ & $9$ & $88.3\%$ & $76.3\%$ & $71.7\%$ & $96.8\%$ & $67.2\%$ & $67.5\%$ & $78.1\%$ & $84.3\%$ & $53.9\%$ & $76.0\%$\\
$2$ & $10$ & $74.9\%$ & $93.6\%$ & $74.7\%$ & $99.5\%$ & $71.2\%$ & $65.1\%$ & $85.6\%$ & $89.1\%$ & $79.7\%$ & $81.5\%$\\
$2$ & $11$ & $87.2\%$ & $88.0\%$ & $68.0\%$ & $77.3\%$ & $85.3\%$ & $79.7\%$ & $77.9\%$ & $81.3\%$ & $79.2\%$ & $80.4\%$\\
$2$ & $12$ & $72.5\%$ & $86.7\%$ & $66.7\%$ & $99.5\%$ & $71.5\%$ & $56.3\%$ & $92.5\%$ & $78.4\%$ & $60.3\%$ & $76.0\%$\\
$2$ & $13$ & $92.0\%$ & $89.9\%$ & $93.3\%$ & $100.0\%$ & $84.0\%$ & $62.4\%$ & $85.3\%$ & $83.2\%$ & $61.6\%$ & $83.5\%$\\
$2$ & $14$ & $93.9\%$ & $99.2\%$ & $91.2\%$ & $100.0\%$ & $82.4\%$ & $80.3\%$ & $97.3\%$ & $93.1\%$ & $88.3\%$ & $91.7\%$\\
$2$ & $15$ & $84.3\%$ & $99.2\%$ & $90.7\%$ & $98.9\%$ & $89.9\%$ & $69.9\%$ & $79.2\%$ & $82.4\%$ & $87.2\%$ & $86.8\%$\\
$2$ & $16$ & $68.5\%$ & $80.5\%$ & $65.9\%$ & $100.0\%$ & $75.2\%$ & $66.4\%$ & $74.1\%$ & $78.1\%$ & $64.8\%$ & $74.8\%$\\
$2$ & $17$ & $93.1\%$ & $89.9\%$ & $82.1\%$ & $100.0\%$ & $72.5\%$ & $78.1\%$ & $91.5\%$ & $88.3\%$ & $89.3\%$ & $87.2\%$\\
$2$ & $18$ & $69.6\%$ & $80.8\%$ & $50.7\%$ & $99.7\%$ & $72.5\%$ & $56.0\%$ & $78.7\%$ & $80.5\%$ & $62.7\%$ & $72.4\%$\\
$2$ & $19$ & $94.9\%$ & $96.5\%$ & $78.7\%$ & $100.0\%$ & $75.7\%$ & $78.4\%$ & $91.7\%$ & $86.9\%$ & $62.7\%$ & $85.1\%$\\ \hline

$3$ & $1$ & $71.5\%$ & $80.3\%$ & $66.9\%$ & $99.7\%$ & $76.0\%$ & $56.8\%$ & $74.1\%$ & $78.7\%$ & $67.7\%$ & $74.6\%$\\
$3$ & $2$ & $94.7\%$ & $98.4\%$ & $86.9\%$ & $70.9\%$ & $64.5\%$ & $57.6\%$ & $93.3\%$ & $56.8\%$ & $56.0\%$ & $75.5\%$\\
$3$ & $3$ & $70.7\%$ & $78.4\%$ & $69.1\%$ & $100.0\%$ & $70.1\%$ & $56.8\%$ & $78.1\%$ & $75.5\%$ & $65.9\%$ & $73.8\%$\\
$3$ & $4$ & $94.9\%$ & $89.3\%$ & $82.7\%$ & $100.0\%$ & $88.8\%$ & $78.4\%$ & $95.2\%$ & $82.1\%$ & $80.5\%$ & $88.0\%$\\
$3$ & $5$ & $94.4\%$ & $99.7\%$ & $90.7\%$ & $100.0\%$ & $100.0\%$ & $74.7\%$ & $96.8\%$ & $96.0\%$ & $86.9\%$ & $93.2\%$\\
$3$ & $6$ & $96.5\%$ & $100.0\%$ & $100.0\%$ & $100.0\%$ & $99.2\%$ & $69.3\%$ & $97.3\%$ & $99.5\%$ & $93.3\%$ & $95.0\%$\\
$3$ & $7$ & $94.4\%$ & $89.1\%$ & $82.1\%$ & $100.0\%$ & $72.5\%$ & $75.2\%$ & $91.2\%$ & $77.6\%$ & $62.7\%$ & $82.8\%$\\
$3$ & $8$ & $94.4\%$ & $88.5\%$ & $91.5\%$ & $100.0\%$ & $73.9\%$ & $84.0\%$ & $92.8\%$ & $91.7\%$ & $61.9\%$ & $86.5\%$\\
$3$ & $9$ & $88.3\%$ & $76.3\%$ & $71.7\%$ & $96.8\%$ & $67.2\%$ & $67.5\%$ & $78.1\%$ & $84.3\%$ & $53.9\%$ & $76.0\%$\\
$3$ & $10$ & $74.9\%$ & $93.6\%$ & $74.7\%$ & $99.5\%$ & $71.2\%$ & $65.1\%$ & $85.6\%$ & $89.1\%$ & $79.7\%$ & $81.5\%$\\
$3$ & $11$ & $87.2\%$ & $88.0\%$ & $68.0\%$ & $77.3\%$ & $85.3\%$ & $79.7\%$ & $77.9\%$ & $81.3\%$ & $79.2\%$ & $80.4\%$\\
$3$ & $12$ & $72.5\%$ & $86.7\%$ & $66.7\%$ & $99.5\%$ & $71.5\%$ & $56.3\%$ & $92.5\%$ & $78.4\%$ & $60.3\%$ & $76.0\%$\\
$3$ & $13$ & $92.0\%$ & $89.9\%$ & $93.3\%$ & $100.0\%$ & $84.0\%$ & $62.4\%$ & $85.3\%$ & $83.2\%$ & $61.6\%$ & $83.5\%$\\
$3$ & $14$ & $93.9\%$ & $99.2\%$ & $91.2\%$ & $100.0\%$ & $82.4\%$ & $80.3\%$ & $97.3\%$ & $93.1\%$ & $88.3\%$ & $91.7\%$\\
$3$ & $15$ & $84.3\%$ & $99.2\%$ & $90.7\%$ & $98.9\%$ & $89.9\%$ & $69.9\%$ & $79.2\%$ & $82.4\%$ & $87.2\%$ & $86.8\%$\\
$3$ & $16$ & $68.5\%$ & $80.5\%$ & $65.9\%$ & $100.0\%$ & $75.2\%$ & $66.4\%$ & $74.1\%$ & $78.1\%$ & $64.8\%$ & $74.8\%$\\
$3$ & $17$ & $93.1\%$ & $89.9\%$ & $82.1\%$ & $100.0\%$ & $72.5\%$ & $78.1\%$ & $91.5\%$ & $88.3\%$ & $89.3\%$ & $87.2\%$\\
$3$ & $18$ & $74.9\%$ & $80.8\%$ & $50.7\%$ & $99.7\%$ & $72.5\%$ & $56.0\%$ & $78.7\%$ & $80.5\%$ & $62.7\%$ & $72.9\%$\\
$3$ & $19$ & $94.9\%$ & $96.5\%$ & $78.7\%$ & $100.0\%$ & $75.7\%$ & $78.4\%$ & $91.7\%$ & $86.9\%$ & $62.7\%$ & $85.1\%$\\

		    \hline
    \end{tabularx}
\end{table*}

\begin{table*}[h!]
	\caption{Validation results --- Adam optimizer, L2 Loss, 2 layers}
	\label{tab:val_2l}
	\begin{tabularx}{0.84\textwidth}{@{}ll|lllllllll|l@{}}
    \hline
	\multicolumn{2}{c|}{\textbf{Setup}} & 
	\multicolumn{10}{c}{\textbf{Performance}} \\
	\hline
	Run & Circuit & 1a & 1b & 1c & 2a & 2b & 2c & 3a & 3b & 3c & Avg. \\
	\hline 

$1$ & $1$ & $74.7\%$ & $84.3\%$ & $70.1\%$ & $100.0\%$ & $76.8\%$ & $53.6\%$ & $92.0\%$ & $80.3\%$ & $57.3\%$ & $76.6\%$\\
$1$ & $2$ & $85.3\%$ & $86.1\%$ & $69.3\%$ & $87.5\%$ & $74.9\%$ & $80.5\%$ & $88.5\%$ & $77.3\%$ & $81.1\%$ & $81.2\%$\\
$1$ & $3$ & $93.1\%$ & $86.9\%$ & $83.7\%$ & $100.0\%$ & $74.4\%$ & $54.7\%$ & $92.3\%$ & $83.7\%$ & $66.9\%$ & $81.7\%$\\
$1$ & $4$ & $94.9\%$ & $96.5\%$ & $80.8\%$ & $100.0\%$ & $98.9\%$ & $90.9\%$ & $95.7\%$ & $91.2\%$ & $76.3\%$ & $91.7\%$\\
$1$ & $5$ & $93.9\%$ & $100.0\%$ & $84.3\%$ & $100.0\%$ & $100.0\%$ & $99.5\%$ & $97.6\%$ & $99.2\%$ & $96.5\%$ & $96.8\%$\\
$1$ & $6$ & $95.2\%$ & $99.5\%$ & $99.7\%$ & $100.0\%$ & $99.7\%$ & $98.1\%$ & $98.9\%$ & $96.0\%$ & $92.3\%$ & $97.7\%$\\
$1$ & $7$ & $92.8\%$ & $99.7\%$ & $81.6\%$ & $100.0\%$ & $80.5\%$ & $95.5\%$ & $92.3\%$ & $86.1\%$ & $86.7\%$ & $90.6\%$\\
$1$ & $8$ & $95.7\%$ & $99.7\%$ & $97.6\%$ & $100.0\%$ & $96.3\%$ & $83.5\%$ & $97.3\%$ & $97.9\%$ & $88.5\%$ & $95.2\%$\\
$1$ & $9$ & $92.0\%$ & $81.3\%$ & $84.8\%$ & $98.7\%$ & $64.8\%$ & $57.9\%$ & $78.1\%$ & $79.5\%$ & $55.2\%$ & $76.9\%$\\
$1$ & $10$ & $83.7\%$ & $95.7\%$ & $74.9\%$ & $100.0\%$ & $81.9\%$ & $69.6\%$ & $93.1\%$ & $85.6\%$ & $76.0\%$ & $84.5\%$\\
$1$ & $11$ & $96.8\%$ & $99.2\%$ & $99.2\%$ & $99.7\%$ & $98.9\%$ & $80.0\%$ & $98.7\%$ & $95.5\%$ & $90.4\%$ & $95.4\%$\\
$1$ & $12$ & $96.0\%$ & $100.0\%$ & $96.8\%$ & $99.7\%$ & $94.1\%$ & $70.9\%$ & $93.9\%$ & $95.2\%$ & $86.4\%$ & $92.6\%$\\
$1$ & $13$ & $92.3\%$ & $100.0\%$ & $97.9\%$ & $100.0\%$ & $98.4\%$ & $96.0\%$ & $98.4\%$ & $94.4\%$ & $93.1\%$ & $96.7\%$\\
$1$ & $14$ & $96.3\%$ & $100.0\%$ & $98.7\%$ & $100.0\%$ & $99.5\%$ & $92.3\%$ & $97.1\%$ & $97.1\%$ & $94.1\%$ & $97.2\%$\\
$1$ & $15$ & $93.6\%$ & $99.7\%$ & $98.1\%$ & $100.0\%$ & $98.9\%$ & $89.3\%$ & $92.5\%$ & $96.3\%$ & $89.6\%$ & $95.3\%$\\
$1$ & $16$ & $94.9\%$ & $99.2\%$ & $87.2\%$ & $100.0\%$ & $83.7\%$ & $58.7\%$ & $91.5\%$ & $80.3\%$ & $59.7\%$ & $83.9\%$\\
$1$ & $17$ & $94.1\%$ & $99.7\%$ & $97.1\%$ & $100.0\%$ & $85.6\%$ & $84.0\%$ & $93.3\%$ & $78.9\%$ & $66.1\%$ & $88.8\%$\\
$1$ & $18$ & $96.5\%$ & $93.1\%$ & $79.5\%$ & $100.0\%$ & $69.1\%$ & $72.8\%$ & $92.5\%$ & $85.3\%$ & $66.4\%$ & $83.9\%$\\
$1$ & $19$ & $94.9\%$ & $99.7\%$ & $97.9\%$ & $100.0\%$ & $99.7\%$ & $96.8\%$ & $97.6\%$ & $96.5\%$ & $90.9\%$ & $97.1\%$\\ \hline

$2$ & $1$ & $74.7\%$ & $84.3\%$ & $70.1\%$ & $100.0\%$ & $76.8\%$ & $53.6\%$ & $92.0\%$ & $80.3\%$ & $57.3\%$ & $76.6\%$\\
$2$ & $2$ & $85.3\%$ & $86.1\%$ & $69.3\%$ & $87.5\%$ & $74.9\%$ & $80.5\%$ & $88.5\%$ & $77.3\%$ & $81.1\%$ & $81.2\%$\\
$2$ & $3$ & $93.1\%$ & $86.9\%$ & $83.7\%$ & $100.0\%$ & $74.4\%$ & $54.7\%$ & $92.3\%$ & $83.7\%$ & $66.9\%$ & $81.7\%$\\
$2$ & $4$ & $94.9\%$ & $96.5\%$ & $80.8\%$ & $100.0\%$ & $98.9\%$ & $90.9\%$ & $95.7\%$ & $91.2\%$ & $76.3\%$ & $91.7\%$\\
$2$ & $5$ & $93.9\%$ & $100.0\%$ & $84.3\%$ & $100.0\%$ & $100.0\%$ & $99.5\%$ & $97.6\%$ & $99.2\%$ & $96.5\%$ & $96.8\%$\\
$2$ & $6$ & $95.2\%$ & $99.5\%$ & $99.7\%$ & $100.0\%$ & $99.7\%$ & $98.1\%$ & $98.9\%$ & $96.0\%$ & $92.3\%$ & $97.7\%$\\
$2$ & $7$ & $92.8\%$ & $99.7\%$ & $81.6\%$ & $100.0\%$ & $80.5\%$ & $95.5\%$ & $92.3\%$ & $86.1\%$ & $86.7\%$ & $90.6\%$\\
$2$ & $8$ & $95.7\%$ & $99.7\%$ & $97.6\%$ & $100.0\%$ & $96.3\%$ & $83.5\%$ & $97.3\%$ & $97.9\%$ & $88.5\%$ & $95.2\%$\\
$2$ & $9$ & $92.0\%$ & $81.3\%$ & $84.8\%$ & $98.7\%$ & $64.8\%$ & $57.9\%$ & $78.1\%$ & $79.5\%$ & $55.2\%$ & $76.9\%$\\
$2$ & $10$ & $83.7\%$ & $95.7\%$ & $74.9\%$ & $100.0\%$ & $81.9\%$ & $69.6\%$ & $93.1\%$ & $85.6\%$ & $76.0\%$ & $84.5\%$\\
$2$ & $11$ & $96.8\%$ & $99.2\%$ & $99.2\%$ & $99.7\%$ & $98.9\%$ & $80.0\%$ & $98.7\%$ & $95.5\%$ & $90.4\%$ & $95.4\%$\\
$2$ & $12$ & $96.0\%$ & $100.0\%$ & $96.8\%$ & $99.7\%$ & $94.1\%$ & $70.9\%$ & $93.9\%$ & $95.2\%$ & $86.4\%$ & $92.6\%$\\
$2$ & $13$ & $92.3\%$ & $100.0\%$ & $97.9\%$ & $100.0\%$ & $98.4\%$ & $96.0\%$ & $98.4\%$ & $94.4\%$ & $93.1\%$ & $96.7\%$\\
$2$ & $14$ & $95.5\%$ & $100.0\%$ & $100.0\%$ & $100.0\%$ & $99.5\%$ & $97.3\%$ & $96.8\%$ & $94.9\%$ & $96.3\%$ & $97.8\%$\\
$2$ & $15$ & $93.6\%$ & $99.7\%$ & $98.1\%$ & $100.0\%$ & $98.9\%$ & $89.3\%$ & $92.5\%$ & $96.3\%$ & $89.6\%$ & $95.3\%$\\
$2$ & $16$ & $94.9\%$ & $99.2\%$ & $87.2\%$ & $100.0\%$ & $83.7\%$ & $58.7\%$ & $91.5\%$ & $80.3\%$ & $59.7\%$ & $83.9\%$\\
$2$ & $17$ & $94.1\%$ & $99.7\%$ & $97.1\%$ & $100.0\%$ & $85.6\%$ & $84.0\%$ & $93.3\%$ & $78.9\%$ & $66.1\%$ & $88.8\%$\\
$2$ & $18$ & $96.5\%$ & $93.1\%$ & $79.5\%$ & $100.0\%$ & $69.1\%$ & $72.8\%$ & $92.5\%$ & $85.3\%$ & $66.4\%$ & $83.9\%$\\
$2$ & $19$ & $94.9\%$ & $99.7\%$ & $97.9\%$ & $100.0\%$ & $99.7\%$ & $96.8\%$ & $97.6\%$ & $96.5\%$ & $90.9\%$ & $97.1\%$\\ \hline

$3$ & $1$ & $74.7\%$ & $84.3\%$ & $70.1\%$ & $100.0\%$ & $76.8\%$ & $53.6\%$ & $92.0\%$ & $80.3\%$ & $57.3\%$ & $76.6\%$\\
$3$ & $2$ & $85.3\%$ & $86.1\%$ & $69.3\%$ & $87.5\%$ & $74.9\%$ & $80.5\%$ & $88.5\%$ & $77.3\%$ & $81.1\%$ & $81.2\%$\\
$3$ & $3$ & $93.1\%$ & $86.9\%$ & $83.7\%$ & $100.0\%$ & $74.4\%$ & $54.7\%$ & $92.3\%$ & $83.7\%$ & $66.9\%$ & $81.7\%$\\
$3$ & $4$ & $94.9\%$ & $96.5\%$ & $80.8\%$ & $100.0\%$ & $98.9\%$ & $90.9\%$ & $95.7\%$ & $91.2\%$ & $76.3\%$ & $91.7\%$\\
$3$ & $5$ & $93.9\%$ & $100.0\%$ & $84.3\%$ & $100.0\%$ & $100.0\%$ & $99.5\%$ & $97.6\%$ & $99.2\%$ & $96.5\%$ & $96.8\%$\\
$3$ & $6$ & $95.2\%$ & $99.5\%$ & $99.7\%$ & $100.0\%$ & $99.7\%$ & $98.1\%$ & $98.9\%$ & $96.0\%$ & $92.3\%$ & $97.7\%$\\
$3$ & $7$ & $92.8\%$ & $99.7\%$ & $81.6\%$ & $100.0\%$ & $80.5\%$ & $95.5\%$ & $92.3\%$ & $86.1\%$ & $86.7\%$ & $90.6\%$\\
$3$ & $8$ & $95.7\%$ & $99.7\%$ & $97.6\%$ & $100.0\%$ & $96.3\%$ & $83.5\%$ & $97.3\%$ & $97.9\%$ & $88.5\%$ & $95.2\%$\\
$3$ & $9$ & $92.0\%$ & $81.3\%$ & $84.8\%$ & $98.7\%$ & $64.8\%$ & $57.9\%$ & $78.1\%$ & $79.5\%$ & $55.2\%$ & $76.9\%$\\
$3$ & $10$ & $83.7\%$ & $95.7\%$ & $74.9\%$ & $100.0\%$ & $81.9\%$ & $69.6\%$ & $93.1\%$ & $85.6\%$ & $76.0\%$ & $84.5\%$\\
$3$ & $11$ & $96.8\%$ & $99.2\%$ & $99.2\%$ & $99.7\%$ & $98.9\%$ & $80.0\%$ & $98.7\%$ & $95.5\%$ & $90.4\%$ & $95.4\%$\\
$3$ & $12$ & $96.0\%$ & $100.0\%$ & $96.8\%$ & $99.7\%$ & $94.1\%$ & $70.9\%$ & $93.9\%$ & $95.2\%$ & $86.4\%$ & $92.6\%$\\
$3$ & $13$ & $92.3\%$ & $100.0\%$ & $97.9\%$ & $100.0\%$ & $98.4\%$ & $96.0\%$ & $98.4\%$ & $94.4\%$ & $93.1\%$ & $96.7\%$\\
$3$ & $14$ & $94.1\%$ & $100.0\%$ & $98.4\%$ & $100.0\%$ & $99.2\%$ & $89.3\%$ & $99.2\%$ & $97.1\%$ & $92.8\%$ & $96.7\%$\\
$3$ & $15$ & $93.6\%$ & $99.7\%$ & $98.1\%$ & $100.0\%$ & $98.9\%$ & $89.3\%$ & $92.5\%$ & $96.3\%$ & $89.6\%$ & $95.3\%$\\
$3$ & $16$ & $94.9\%$ & $99.2\%$ & $87.2\%$ & $100.0\%$ & $83.7\%$ & $58.7\%$ & $91.5\%$ & $80.3\%$ & $59.7\%$ & $83.9\%$\\
$3$ & $17$ & $94.1\%$ & $99.7\%$ & $97.1\%$ & $100.0\%$ & $85.6\%$ & $84.0\%$ & $93.3\%$ & $78.9\%$ & $66.1\%$ & $88.8\%$\\
$3$ & $19$ & $94.9\%$ & $99.7\%$ & $97.9\%$ & $100.0\%$ & $99.7\%$ & $96.8\%$ & $97.6\%$ & $96.5\%$ & $90.9\%$ & $97.1\%$\\ 
	    \hline
    \end{tabularx}
\end{table*}

\clearpage
\section{\bf{Hyperparameter search results}} \label{sec:app_hyp}

\begin{table*}[h!]	
    \caption{Hyperparameter search results --- Adam optimizer, L1 loss}
	\label{tab:adam_l1}
	\begin{tabularx}{0.84\textwidth}{@{}ll|lllllllll|l@{}}
    \hline
	\multicolumn{2}{c|}{\textbf{Setup}} & 
	\multicolumn{10}{c}{\textbf{Performance}} \\
	\hline
    Layers & Circuit & 1a & 1b & 1c & 2a & 2b & 2c & 3a & 3b & 3c & Avg. \\
	\hline 
	
	$1$ & $1$ & $63.2\%$ & $82.1\%$ & $74.7\%$ & $96.8\%$ & $71.2\%$ & $54.7\%$ & $71.7\%$ & $81.1\%$ & $51.2\%$ & $71.9\%$\\
    $1$ & $2$ & $92.5\%$ & $77.9\%$ & $70.1\%$ & $48.8\%$ & $53.9\%$ & $52.8\%$ & $79.2\%$ & $46.4\%$ & $49.9\%$ & $63.5\%$\\
    $1$ & $3$ & $76.0\%$ & $77.6\%$ & $73.1\%$ & $98.4\%$ & $72.8\%$ & $54.9\%$ & $72.3\%$ & $77.9\%$ & $57.6\%$ & $73.4\%$\\
    $1$ & $4$ & $93.6\%$ & $90.7\%$ & $81.3\%$ & $51.7\%$ & $83.5\%$ & $84.0\%$ & $86.9\%$ & $77.9\%$ & $76.0\%$ & $80.6\%$\\
    $1$ & $5$ & $90.7\%$ & $95.7\%$ & $93.9\%$ & $90.7\%$ & $82.4\%$ & $49.3\%$ & $94.4\%$ & $89.9\%$ & $77.1\%$ & $84.9\%$\\
    $1$ & $6$ & $87.2\%$ & $91.5\%$ & $89.6\%$ & $97.9\%$ & $91.7\%$ & $93.6\%$ & $93.3\%$ & $86.1\%$ & $86.1\%$ & $90.8\%$\\
    $1$ & $7$ & $88.5\%$ & $86.9\%$ & $79.5\%$ & $99.7\%$ & $66.7\%$ & $59.5\%$ & $83.2\%$ & $79.5\%$ & $80.3\%$ & $80.4\%$\\
    $1$ & $8$ & $93.3\%$ & $92.0\%$ & $88.5\%$ & $98.4\%$ & $87.5\%$ & $76.8\%$ & $82.4\%$ & $82.7\%$ & $71.2\%$ & $85.9\%$\\
    $1$ & $9$ & $78.4\%$ & $62.1\%$ & $85.6\%$ & $86.1\%$ & $71.2\%$ & $57.1\%$ & $74.4\%$ & $81.9\%$ & $56.3\%$ & $72.6\%$\\
    $1$ & $11$ & $69.1\%$ & $78.7\%$ & $53.9\%$ & $80.0\%$ & $83.7\%$ & $73.9\%$ & $83.5\%$ & $80.5\%$ & $76.0\%$ & $75.5\%$\\
    $1$ & $12$ &$77.1\%$ & $87.7\%$ & $63.7\%$ & $91.5\%$ & $82.4\%$ & $56.8\%$ & $67.2\%$ & $83.5\%$ & $46.7\%$ & $72.9\%$\\
    $1$ & $13$ & $93.6\%$ & $91.2\%$ & $84.3\%$ & $97.9\%$ & $84.5\%$ & $73.3\%$ & $86.7\%$ & $81.1\%$ & $78.1\%$ & $85.6\%$\\
    $1$ & $14$ & $89.6\%$ & $97.6\%$ & $89.3\%$ & $92.8\%$ & $94.1\%$ & $73.9\%$ & $89.6\%$ & $90.7\%$ & $77.3\%$ & $88.3\%$\\
    $1$ & $15$ & $87.2\%$ & $90.7\%$ & $79.5\%$ & $84.8\%$ & $77.3\%$ & $63.2\%$ & $84.0\%$ & $88.0\%$ & $81.9\%$ & $81.8\%$\\
    $1$ & $16$ & $62.7\%$ & $82.1\%$ & $66.4\%$ & $100.0\%$ & $72.3\%$ & $55.7\%$ & $74.4\%$ & $77.6\%$ & $48.8\%$ & $71.1\%$\\
    $1$ & $17$ & $94.4\%$ & $76.3\%$ & $83.7\%$ & $95.7\%$ & $79.7\%$ & $63.5\%$ & $85.9\%$ & $82.7\%$ & $79.5\%$ & $82.4\%$\\
    $1$ & $18$ & $73.9\%$ & $82.4\%$ & $61.6\%$ & $97.9\%$ & $70.9\%$ & $61.1\%$ & $81.3\%$ & $84.0\%$ & $48.8\%$ & $73.5\%$\\
    $1$ & $19$ & $74.9\%$ & $90.4\%$ & $83.5\%$ & $99.7\%$ & $70.1\%$ & $64.8\%$ & $89.6\%$ & $83.7\%$ & $63.7\%$ & $80.1\%$\\ \hline
	
	$2$ & $1$ & $63.5\%$ & $70.1\%$ & $62.9\%$ & $94.4\%$ & $72.3\%$ & $57.1\%$ & $84.0\%$ & $80.8\%$ & $50.7\%$ & $70.6\%$\\
	$2$ & $2$ & $64.5\%$ & $89.6\%$ & $82.7\%$ & $87.2\%$ & $86.4\%$ & $80.8\%$ & $80.5\%$ & $77.3\%$ & $71.7\%$ & $80.1\%$\\
	$2$ & $3$ & $90.7\%$ & $76.3\%$ & $85.6\%$ & $87.7\%$ & $84.5\%$ & $49.9\%$ & $79.7\%$ & $80.8\%$ & $72.5\%$ & $78.6\%$\\
	$2$ & $4$ & $88.3\%$ & $91.2\%$ & $93.1\%$ & $89.9\%$ & $85.3\%$ & $70.4\%$ & $87.5\%$ & $88.8\%$ & $66.1\%$ & $84.5\%$\\
	$2$ & $5$ & $93.6\%$ & $93.9\%$ & $88.3\%$ & $93.9\%$ & $89.3\%$ & $93.6\%$ & $90.1\%$ & $90.7\%$ & $84.0\%$ & $90.8\%$\\
	$2$ & $6$ & $92.8\%$ & $90.4\%$ & $89.3\%$ & $99.7\%$ & $98.1\%$ & $85.6\%$ & $90.7\%$ & $88.8\%$ & $80.8\%$ & $90.7\%$\\
	$2$ & $7$ & $90.1\%$ & $90.7\%$ & $89.3\%$ & $94.9\%$ & $72.0\%$ & $64.8\%$ & $90.4\%$ & $84.8\%$ & $78.1\%$ & $83.9\%$\\
	$2$ & $8$ & $93.6\%$ & $97.6\%$ & $85.6\%$ & $99.2\%$ & $88.3\%$ & $68.8\%$ & $88.5\%$ & $84.8\%$ & $79.2\%$ & $87.3\%$\\
	$2$ & $9$ & $85.3\%$ & $61.3\%$ & $86.4\%$ & $93.3\%$ & $72.8\%$ & $67.7\%$ & $85.1\%$ & $72.0\%$ & $66.4\%$ & $76.7\%$\\
	$2$ & $11$ & $87.7\%$ & $93.6\%$ & $86.1\%$ & $95.7\%$ & $89.6\%$ & $83.7\%$ & $85.6\%$ & $85.9\%$ & $88.0\%$ & $88.4\%$\\
	$2$ & $12$ & $89.6\%$ & $93.3\%$ & $84.3\%$ & $97.3\%$ & $82.1\%$ & $82.4\%$ & $88.0\%$ & $75.5\%$ & $78.4\%$ & $85.7\%$\\
	$2$ & $13$ & $86.1\%$ & $98.9\%$ & $87.2\%$ & $99.7\%$ & $91.5\%$ & $89.1\%$ & $91.7\%$ & $73.1\%$ & $82.1\%$ & $88.8\%$\\
	$2$ & $14$ & $92.0\%$ & $85.1\%$ & $92.3\%$ & $100.0\%$ & $92.3\%$ & $81.1\%$ & $86.9\%$ & $88.8\%$ & $86.4\%$ & $89.4\%$\\
	$2$ & $15$ & $79.7\%$ & $92.0\%$ & $93.9\%$ & $97.6\%$ & $84.3\%$ & $80.5\%$ & $83.7\%$ & $82.7\%$ & $76.8\%$ & $85.7\%$\\
	$2$ & $16$ & $85.1\%$ & $73.6\%$ & $82.7\%$ & $97.3\%$ & $85.1\%$ & $71.7\%$ & $86.1\%$ & $78.9\%$ & $58.9\%$ & $79.9\%$\\
	$2$ & $17$ & $93.6\%$ & $93.3\%$ & $80.8\%$ & $98.4\%$ & $88.8\%$ & $74.1\%$ & $92.5\%$ & $92.8\%$ & $64.8\%$ & $86.6\%$\\
	$2$ & $18$ & $91.2\%$ & $88.8\%$ & $86.9\%$ & $99.5\%$ & $87.5\%$ & $56.8\%$ & $85.1\%$ & $82.7\%$ & $73.3\%$ & $83.5\%$\\
	$2$ & $19$ & $89.6\%$ & $96.0\%$ & $89.3\%$ & $98.4\%$ & $92.8\%$ & $83.7\%$ & $88.5\%$ & $92.0\%$ & $83.2\%$ & $90.4\%$\\
	    \hline
    \end{tabularx}
\end{table*}


\begin{table*}[h!]
	\caption{Hyperparameter search results --- Adam optimizer, L2 loss}
	\label{tab:adam_l2}
	\begin{tabularx}{0.86\textwidth}{@{}ll|lllllllll|l@{}}
    \hline
	\multicolumn{2}{c|}{\textbf{Setup}} & 
	\multicolumn{10}{c}{\textbf{Performance}} \\
	\hline
	Layers & Circuit & 1a & 1b & 1c & 2a & 2b & 2c & 3a & 3b & 3c & Avg. \\
	\hline 
	
	$1$ & $1$ & $74.7\%$ & $77.3\%$ & $65.1\%$ & $100.0\%$ & $75.2\%$ & $55.2\%$ & $76.3\%$ & $76.0\%$ & $66.4\%$ & $74.0\%$\\
    $1$ & $2$ & $95.5\%$ & $99.2\%$ & $83.5\%$ & $72.0\%$ & $57.6\%$ & $59.2\%$ & $92.5\%$ & $57.1\%$ & $65.6\%$ & $75.8\%$\\
    $1$ & $3$ & $68.8\%$ & $80.0\%$ & $51.2\%$ & $100.0\%$ & $71.7\%$ & $57.1\%$ & $90.9\%$ & $80.5\%$ & $64.8\%$ & $73.9\%$\\
    $1$ & $4$ & $95.5\%$ & $91.7\%$ & $85.1\%$ & $100.0\%$ & $74.7\%$ & $70.1\%$ & $88.0\%$ & $89.1\%$ & $64.0\%$ & $84.2\%$\\
    $1$ & $5$ & $96.0\%$ & $99.2\%$ & $93.3\%$ & $100.0\%$ & $87.2\%$ & $88.0\%$ & $98.4\%$ & $89.3\%$ & $89.1\%$ & $93.4\%$\\
    $1$ & $6$ & $96.0\%$ & $100.0\%$ & $97.1\%$ & $100.0\%$ & $100.0\%$ & $78.9\%$ & $97.9\%$ & $97.1\%$ & $97.1\%$ & $96.0\%$\\
    $1$ & $7$ & $96.8\%$ & $85.9\%$ & $82.4\%$ & $100.0\%$ & $70.4\%$ & $68.8\%$ & $94.1\%$ & $80.5\%$ & $65.6\%$ & $82.7\%$\\
    $1$ & $8$ & $96.5\%$ & $97.6\%$ & $82.7\%$ & $99.7\%$ & $71.5\%$ & $82.9\%$ & $93.6\%$ & $87.5\%$ & $65.6\%$ & $86.4\%$\\
    $1$ & $9$ & $86.4\%$ & $80.3\%$ & $90.4\%$ & $97.1\%$ & $64.3\%$ & $55.2\%$ & $78.9\%$ & $83.5\%$ & $54.9\%$ & $76.8\%$\\
    $1$ & $11$ & $82.7\%$ & $82.4\%$ & $57.6\%$ & $93.6\%$ & $83.5\%$ & $84.5\%$ & $94.1\%$ & $81.6\%$ & $80.3\%$ & $82.3\%$\\
    $1$ & $12$ & $78.4\%$ & $81.9\%$ & $70.7\%$ & $100.0\%$ & $80.0\%$ & $57.1\%$ & $92.3\%$ & $80.0\%$ & $59.5\%$ & $77.7\%$\\
    $1$ & $13$ & $95.2\%$ & $85.1\%$ & $92.3\%$ & $100.0\%$ & $81.6\%$ & $62.9\%$ & $94.4\%$ & $80.5\%$ & $69.1\%$ & $84.6\%$\\
    $1$ & $14$ & $95.2\%$ & $100.0\%$ & $93.1\%$ & $100.0\%$ & $98.9\%$ & $77.1\%$ & $93.3\%$ & $85.3\%$ & $91.7\%$ & $92.7\%$\\
    $1$ & $15$ & $72.5\%$ & $86.1\%$ & $93.9\%$ & $100.0\%$ & $68.3\%$ & $65.9\%$ & $87.7\%$ & $91.5\%$ & $78.7\%$ & $82.7\%$\\
    $1$ & $16$ & $70.9\%$ & $78.1\%$ & $63.7\%$ & $100.0\%$ & $70.1\%$ & $53.3\%$ & $78.9\%$ & $81.1\%$ & $67.2\%$ & $73.7\%$\\
    $1$ & $17$ & $93.3\%$ & $98.4\%$ & $95.7\%$ & $100.0\%$ & $84.5\%$ & $86.1\%$ & $83.2\%$ & $83.7\%$ & $70.1\%$ & $88.4\%$\\
    $1$ & $18$ & $69.6\%$ & $74.9\%$ & $54.4\%$ & $100.0\%$ & $73.6\%$ & $61.1\%$ & $77.3\%$ & $81.9\%$ & $56.5\%$ & $72.1\%$\\
    $1$ & $19$ & $95.7\%$ & $86.9\%$ & $78.1\%$ & $100.0\%$ & $85.1\%$ & $56.5\%$ & $89.9\%$ & $86.7\%$ & $62.4\%$ & $82.4\%$\\ \hline
    
    $2$ & $1$ & $83.7\%$ & $83.2\%$ & $68.3\%$ & $100.0\%$ & $69.3\%$ & $57.1\%$ & $78.1\%$ & $80.5\%$ & $65.1\%$ & $76.1\%$\\
    $2$ & $2$ & $89.3\%$ & $91.2\%$ & $98.7\%$ & $80.0\%$ & $85.1\%$ & $82.9\%$ & $88.8\%$ & $82.9\%$ & $65.3\%$ & $84.9\%$\\
    $2$ & $3$ & $96.8\%$ & $90.1\%$ & $82.9\%$ & $100.0\%$ & $73.3\%$ & $70.4\%$ & $93.6\%$ & $86.7\%$ & $72.8\%$ & $85.2\%$\\
    $2$ & $4$ & $93.9\%$ & $98.4\%$ & $81.9\%$ & $100.0\%$ & $95.5\%$ & $94.4\%$ & $95.2\%$ & $83.5\%$ & $63.7\%$ & $89.6\%$\\
    $2$ & $5$ & $96.0\%$ & $99.7\%$ & $98.9\%$ & $100.0\%$ & $100.0\%$ & $89.9\%$ & $98.1\%$ & $98.1\%$ & $93.9\%$ & $97.2\%$\\
    $2$ & $6$ & $96.3\%$ & $100.0\%$ & $98.1\%$ & $100.0\%$ & $100.0\%$ & $94.1\%$ & $98.9\%$ & $97.3\%$ & $93.1\%$ & $97.5\%$\\
    $2$ & $7$ & $96.0\%$ & $99.2\%$ & $92.3\%$ & $100.0\%$ & $98.7\%$ & $84.5\%$ & $94.1\%$ & $85.6\%$ & $94.7\%$ & $93.9\%$\\
    $2$ & $8$ & $96.5\%$ & $92.8\%$ & $89.9\%$ & $100.0\%$ & $99.5\%$ & $84.5\%$ & $95.2\%$ & $87.5\%$ & $87.5\%$ & $92.6\%$\\
    $2$ & $9$ & $88.0\%$ & $88.5\%$ & $91.2\%$ & $97.1\%$ & $72.3\%$ & $77.1\%$ & $81.9\%$ & $80.3\%$ & $67.5\%$ & $82.6\%$\\
    $2$ & $11$ & $95.5\%$ & $100.0\%$ & $99.5\%$ & $100.0\%$ & $88.3\%$ & $74.1\%$ & $98.1\%$ & $96.5\%$ & $88.8\%$ & $93.4\%$\\
    $2$ & $12$ & $94.1\%$ & $98.1\%$ & $80.5\%$ & $100.0\%$ & $94.9\%$ & $69.6\%$ & $97.1\%$ & $96.3\%$ & $87.5\%$ & $90.9\%$\\
    $2$ & $13$ & $95.7\%$ & $100.0\%$ & $99.5\%$ & $100.0\%$ & $98.7\%$ & $93.9\%$ & $96.5\%$ & $94.9\%$ & $94.1\%$ & $97.0\%$\\
    $2$ & $14$ & $95.7\%$ & $100.0\%$ & $98.1\%$ & $99.7\%$ & $100.0\%$ & $92.3\%$ & $96.8\%$ & $95.7\%$ & $93.1\%$ & $96.8\%$\\
    $2$ & $15$ & $94.7\%$ & $97.9\%$ & $98.1\%$ & $100.0\%$ & $98.9\%$ & $73.1\%$ & $97.3\%$ & $96.0\%$ & $87.2\%$ & $93.7\%$\\
    $2$ & $16$ & $95.5\%$ & $89.6\%$ & $81.3\%$ & $100.0\%$ & $73.1\%$ & $56.0\%$ & $84.8\%$ & $87.7\%$ & $61.3\%$ & $81.0\%$\\
    $2$ & $17$ & $95.2\%$ & $99.2\%$ & $81.3\%$ & $99.2\%$ & $73.6\%$ & $77.1\%$ & $94.4\%$ & $93.6\%$ & $90.7\%$ & $89.4\%$\\
    $2$ & $18$ & $94.4\%$ & $94.4\%$ & $83.5\%$ & $100.0\%$ & $86.4\%$ & $54.7\%$ & $90.9\%$ & $85.9\%$ & $71.5\%$ & $84.6\%$\\
    $2$ & $19$ & $95.2\%$ & $100.0\%$ & $95.5\%$ & $100.0\%$ & $99.5\%$ & $85.1\%$ & $97.3\%$ & $90.9\%$ & $87.5\%$ & $94.5\%$\\

    \hline
     \end{tabularx}
\end{table*}

\begin{table*}[h!]
	\caption{Hyperparameter search results --- Gradient Decent optimizer, L1 loss}
	\label{tab:gd_l1}
	\begin{tabularx}{0.84\textwidth}{@{}ll|lllllllll|l@{}}
    \hline
	\multicolumn{2}{c|}{\textbf{Setup}} & 
	\multicolumn{10}{c}{\textbf{Performance}} \\
	\hline
	Layers & Circuit & 1a & 1b & 1c & 2a & 2b & 2c & 3a & 3b & 3c & Avg. \\
	\hline 

    $1$ & $1$ & $71.7\%$ & $78.1\%$ & $48.3\%$ & $100.0\%$ & $73.6\%$ & $48.0\%$ & $75.7\%$ & $79.2\%$ & $50.1\%$ & $69.4\%$\\
    $1$ & $2$ & $95.5\%$ & $93.6\%$ & $85.1\%$ & $50.9\%$ & $60.0\%$ & $49.6\%$ & $76.8\%$ & $52.8\%$ & $46.1\%$ & $67.8\%$\\
    $1$ & $3$ & $73.6\%$ & $81.3\%$ & $52.5\%$ & $100.0\%$ & $71.5\%$ & $54.1\%$ & $78.9\%$ & $78.4\%$ & $49.3\%$ & $71.1\%$\\
    $1$ & $4$ & $86.9\%$ & $93.1\%$ & $69.6\%$ & $96.5\%$ & $70.1\%$ & $53.6\%$ & $80.8\%$ & $78.7\%$ & $48.5\%$ & $75.3\%$\\
    $1$ & $5$ & $93.3\%$ & $89.1\%$ & $76.5\%$ & $98.4\%$ & $74.4\%$ & $54.1\%$ & $78.4\%$ & $78.4\%$ & $45.9\%$ & $76.5\%$\\
    $1$ & $6$ & $93.3\%$ & $94.9\%$ & $88.5\%$ & $100.0\%$ & $79.2\%$ & $68.3\%$ & $85.9\%$ & $80.0\%$ & $64.0\%$ & $83.8\%$\\
    $1$ & $7$ & $85.9\%$ & $82.1\%$ & $74.4\%$ & $100.0\%$ & $70.4\%$ & $56.5\%$ & $79.2\%$ & $78.7\%$ & $52.8\%$ & $75.6\%$\\
    $1$ & $8$ & $88.5\%$ & $81.9\%$ & $73.3\%$ & $99.7\%$ & $74.4\%$ & $60.0\%$ & $78.9\%$ & $79.5\%$ & $52.3\%$ & $76.5\%$\\
    $1$ & $9$ & $88.8\%$ & $80.3\%$ & $75.7\%$ & $88.0\%$ & $69.3\%$ & $52.5\%$ & $76.0\%$ & $82.1\%$ & $52.5\%$ & $73.9\%$\\
    $1$ & $11$ & $50.4\%$ & $89.3\%$ & $53.9\%$ & $82.9\%$ & $76.8\%$ & $72.3\%$ & $76.5\%$ & $78.9\%$ & $75.7\%$ & $73.0\%$\\
    $1$ & $12$ & $67.7\%$ & $86.1\%$ & $58.4\%$ & $100.0\%$ & $78.9\%$ & $50.1\%$ & $79.5\%$ & $67.2\%$ & $49.6\%$ & $70.8\%$\\
    $1$ & $13$ & $86.1\%$ & $85.3\%$ & $73.3\%$ & $100.0\%$ & $69.6\%$ & $56.5\%$ & $79.7\%$ & $79.5\%$ & $55.2\%$ & $76.1\%$\\
    $1$ & $14$ & $88.8\%$ & $87.5\%$ & $82.9\%$ & $91.5\%$ & $78.7\%$ & $48.8\%$ & $78.4\%$ & $79.7\%$ & $47.5\%$ & $76.0\%$\\
    $1$ & $15$ & $79.5\%$ & $85.1\%$ & $88.8\%$ & $79.2\%$ & $68.3\%$ & $63.5\%$ & $89.1\%$ & $73.9\%$ & $48.8\%$ & $75.1\%$\\
    $1$ & $16$ & $72.5\%$ & $79.2\%$ & $47.7\%$ & $100.0\%$ & $73.6\%$ & $48.5\%$ & $75.7\%$ & $79.5\%$ & $50.1\%$ & $69.7\%$\\
    $1$ & $17$ & $81.3\%$ & $89.1\%$ & $69.9\%$ & $97.3\%$ & $74.9\%$ & $54.7\%$ & $79.2\%$ & $78.1\%$ & $50.7\%$ & $75.0\%$\\
    $1$ & $18$ & $73.6\%$ & $78.9\%$ & $52.5\%$ & $100.0\%$ & $71.5\%$ & $52.5\%$ & $78.9\%$ & $78.7\%$ & $49.3\%$ & $70.7\%$\\
    $1$ & $19$ & $85.3\%$ & $89.1\%$ & $75.7\%$ & $98.9\%$ & $74.1\%$ & $56.8\%$ & $78.7\%$ & $79.2\%$ & $54.1\%$ & $76.9\%$\\ \hline
    
    $2$ & $1$ & $73.3\%$ & $82.4\%$ & $50.9\%$ & $100.0\%$ & $74.7\%$ & $47.5\%$ & $75.2\%$ & $79.2\%$ & $50.4\%$ & $70.4\%$\\
    $2$ & $2$ & $81.9\%$ & $89.9\%$ & $95.7\%$ & $54.4\%$ & $78.1\%$ & $47.7\%$ & $68.3\%$ & $77.6\%$ & $79.2\%$ & $74.8\%$\\
    $2$ & $3$ & $88.5\%$ & $85.3\%$ & $65.9\%$ & $99.5\%$ & $72.0\%$ & $53.6\%$ & $78.9\%$ & $78.4\%$ & $55.7\%$ & $75.3\%$\\
    $2$ & $4$ & $91.5\%$ & $97.3\%$ & $75.2\%$ & $96.3\%$ & $71.5\%$ & $56.8\%$ & $79.5\%$ & $78.7\%$ & $48.5\%$ & $77.2\%$\\
    $2$ & $5$ & $94.7\%$ & $85.6\%$ & $83.2\%$ & $99.7\%$ & $81.3\%$ & $59.5\%$ & $81.6\%$ & $78.9\%$ & $49.9\%$ & $79.4\%$\\
    $2$ & $6$ & $92.0\%$ & $97.1\%$ & $82.1\%$ & $100.0\%$ & $83.5\%$ & $74.9\%$ & $88.5\%$ & $86.7\%$ & $58.9\%$ & $84.9\%$\\
    $2$ & $7$ & $93.6\%$ & $82.1\%$ & $81.3\%$ & $96.3\%$ & $69.1\%$ & $45.6\%$ & $78.9\%$ & $79.2\%$ & $51.7\%$ & $75.3\%$\\
    $2$ & $8$ & $91.7\%$ & $79.5\%$ & $88.3\%$ & $94.4\%$ & $76.5\%$ & $63.5\%$ & $78.9\%$ & $77.1\%$ & $48.8\%$ & $77.6\%$\\
    $2$ & $9$ & $87.5\%$ & $78.7\%$ & $70.4\%$ & $83.2\%$ & $69.9\%$ & $64.0\%$ & $66.1\%$ & $79.5\%$ & $44.0\%$ & $71.5\%$\\
    $2$ & $11$ & $76.8\%$ & $98.1\%$ & $89.1\%$ & $89.9\%$ & $88.5\%$ & $76.5\%$ & $80.3\%$ & $88.8\%$ & $74.1\%$ & $84.7\%$\\
    $2$ & $12$ & $88.3\%$ & $89.6\%$ & $84.8\%$ & $98.4\%$ & $84.3\%$ & $53.6\%$ & $86.9\%$ & $81.6\%$ & $58.7\%$ & $80.7\%$\\
    $2$ & $13$ & $92.8\%$ & $88.5\%$ & $78.1\%$ & $99.7\%$ & $75.7\%$ & $55.7\%$ & $83.7\%$ & $78.9\%$ & $51.5\%$ & $78.3\%$\\
    $2$ & $14$ & $92.8\%$ & $86.7\%$ & $89.1\%$ & $100.0\%$ & $76.0\%$ & $62.1\%$ & $81.3\%$ & $79.7\%$ & $53.1\%$ & $80.1\%$\\
    $2$ & $15$ & $88.8\%$ & $81.3\%$ & $89.3\%$ & $87.5\%$ & $84.5\%$ & $79.5\%$ & $85.3\%$ & $74.4\%$ & $53.3\%$ & $80.4\%$\\
    $2$ & $16$ & $76.5\%$ & $83.7\%$ & $74.7\%$ & $100.0\%$ & $76.0\%$ & $51.7\%$ & $75.7\%$ & $79.5\%$ & $53.6\%$ & $74.6\%$\\
    $2$ & $17$ & $92.8\%$ & $82.1\%$ & $82.9\%$ & $92.3\%$ & $74.7\%$ & $62.4\%$ & $74.9\%$ & $79.5\%$ & $52.5\%$ & $77.1\%$\\
    $2$ & $18$ & $93.1\%$ & $83.7\%$ & $76.5\%$ & $98.7\%$ & $69.1\%$ & $51.5\%$ & $78.4\%$ & $78.4\%$ & $55.5\%$ & $76.1\%$\\
    $2$ & $19$ & $90.9\%$ & $92.3\%$ & $85.3\%$ & $97.1\%$ & $74.1\%$ & $50.1\%$ & $80.3\%$ & $77.9\%$ & $51.2\%$ & $77.7\%$\\

    \hline
    \end{tabularx}
\end{table*}

\begin{table*}[h!]
	\caption{Hyperparameter search results --- Gradient Decent optimizer, L2 loss}
	\label{tab:gd_l2}
	\begin{tabularx}{0.84\textwidth}{@{}ll|lllllllll|l@{}}
    \hline
	\multicolumn{2}{c}{\textbf{Setup}} & 
	\multicolumn{10}{c}{\textbf{Performance}} \\
	\hline
	Layers & Circuit & 1a & 1b & 1c & 2a & 2b & 2c & 3a & 3b & 3c & Avg. \\
	\hline 
    $1$ & $1$ & $69.1\%$ & $80.5\%$ & $50.9\%$ & $100.0\%$ & $74.9\%$ & $51.7\%$ & $74.4\%$ & $78.4\%$ & $52.0\%$ & $70.2\%$\\
    $1$ & $2$ & $94.1\%$ & $97.9\%$ & $80.0\%$ & $56.0\%$ & $60.0\%$ & $50.7\%$ & $75.5\%$ & $59.2\%$ & $48.0\%$ & $69.0\%$\\
    $1$ & $3$ & $71.7\%$ & $78.4\%$ & $51.2\%$ & $100.0\%$ & $69.6\%$ & $54.4\%$ & $79.5\%$ & $78.4\%$ & $51.2\%$ & $70.5\%$\\
    $1$ & $4$ & $94.7\%$ & $96.3\%$ & $78.7\%$ & $100.0\%$ & $74.1\%$ & $70.7\%$ & $91.2\%$ & $78.7\%$ & $61.1\%$ & $82.8\%$\\
    $1$ & $5$ & $96.0\%$ & $96.8\%$ & $90.7\%$ & $99.7\%$ & $88.3\%$ & $60.8\%$ & $96.0\%$ & $82.1\%$ & $62.9\%$ & $85.9\%$\\
    $1$ & $6$ & $95.2\%$ & $98.1\%$ & $88.5\%$ & $100.0\%$ & $99.2\%$ & $70.7\%$ & $93.9\%$ & $83.7\%$ & $76.8\%$ & $89.6\%$\\
    $1$ & $7$ & $95.5\%$ & $85.1\%$ & $79.7\%$ & $100.0\%$ & $70.9\%$ & $61.9\%$ & $78.9\%$ & $78.9\%$ & $53.9\%$ & $78.3\%$\\
    $1$ & $8$ & $89.6\%$ & $96.5\%$ & $82.1\%$ & $100.0\%$ & $77.3\%$ & $55.7\%$ & $83.2\%$ & $79.5\%$ & $53.3\%$ & $79.7\%$\\
    $1$ & $9$ & $89.3\%$ & $61.1\%$ & $78.9\%$ & $99.7\%$ & $64.5\%$ & $62.7\%$ & $80.0\%$ & $79.5\%$ & $51.7\%$ & $74.2\%$\\
    $1$ & $11$ & $55.2\%$ & $93.6\%$ & $77.1\%$ & $92.3\%$ & $81.3\%$ & $81.9\%$ & $79.7\%$ & $70.4\%$ & $70.7\%$ & $78.0\%$\\
    $1$ & $12$ & $74.1\%$ & $85.9\%$ & $66.7\%$ & $100.0\%$ & $78.4\%$ & $60.3\%$ & $78.9\%$ & $78.4\%$ & $48.8\%$ & $74.6\%$\\
    $1$ & $13$ & $96.3\%$ & $83.7\%$ & $80.5\%$ & $100.0\%$ & $73.1\%$ & $54.9\%$ & $85.9\%$ & $77.9\%$ & $52.3\%$ & $78.3\%$\\
    $1$ & $14$ & $93.6\%$ & $88.5\%$ & $97.9\%$ & $99.7\%$ & $77.3\%$ & $55.5\%$ & $80.0\%$ & $79.5\%$ & $55.2\%$ & $80.8\%$\\
    $1$ & $15$ & $81.1\%$ & $80.3\%$ & $91.5\%$ & $84.5\%$ & $92.3\%$ & $62.9\%$ & $91.5\%$ & $90.9\%$ & $54.7\%$ & $81.1\%$\\
    $1$ & $16$ & $69.1\%$ & $80.5\%$ & $51.2\%$ & $100.0\%$ & $74.9\%$ & $49.1\%$ & $74.4\%$ & $78.4\%$ & $50.4\%$ & $69.8\%$\\
    $1$ & $17$ & $92.5\%$ & $90.9\%$ & $80.8\%$ & $99.5\%$ & $79.2\%$ & $64.5\%$ & $80.8\%$ & $79.5\%$ & $55.7\%$ & $80.4\%$\\
    $1$ & $18$ & $71.7\%$ & $78.7\%$ & $51.2\%$ & $100.0\%$ & $69.6\%$ & $57.1\%$ & $79.5\%$ & $78.4\%$ & $51.2\%$ & $70.8\%$\\
    $1$ & $19$ & $94.7\%$ & $96.8\%$ & $86.1\%$ & $100.0\%$ & $74.9\%$ & $56.3\%$ & $79.7\%$ & $78.9\%$ & $59.5\%$ & $80.8\%$\\ \hline

    $2$ & $1$ & $75.5\%$ & $83.7\%$ & $47.2\%$ & $100.0\%$ & $76.0\%$ & $50.7\%$ & $74.4\%$ & $80.3\%$ & $51.7\%$ & $71.1\%$\\
    $2$ & $2$ & $85.6\%$ & $91.2\%$ & $97.9\%$ & $88.5\%$ & $87.2\%$ & $81.1\%$ & $67.5\%$ & $74.1\%$ & $81.6\%$ & $83.9\%$\\
    $2$ & $3$ & $93.6\%$ & $86.9\%$ & $80.5\%$ & $100.0\%$ & $70.9\%$ & $62.4\%$ & $80.8\%$ & $78.9\%$ & $63.7\%$ & $79.8\%$\\
    $2$ & $4$ & $96.0\%$ & $99.5\%$ & $84.3\%$ & $99.7\%$ & $71.2\%$ & $55.7\%$ & $87.2\%$ & $78.1\%$ & $59.5\%$ & $81.2\%$\\
    $2$ & $5$ & $96.0\%$ & $99.7\%$ & $97.1\%$ & $100.0\%$ & $96.0\%$ & $81.1\%$ & $95.5\%$ & $92.8\%$ & $78.4\%$ & $92.9\%$\\
    $2$ & $6$ & $94.9\%$ & $98.9\%$ & $97.3\%$ & $100.0\%$ & $99.7\%$ & $81.3\%$ & $96.8\%$ & $82.9\%$ & $86.4\%$ & $93.2\%$\\
    $2$ & $7$ & $96.0\%$ & $91.5\%$ & $81.1\%$ & $98.9\%$ & $72.3\%$ & $61.3\%$ & $92.8\%$ & $79.2\%$ & $53.1\%$ & $80.7\%$\\
    $2$ & $8$ & $93.9\%$ & $86.1\%$ & $88.0\%$ & $100.0\%$ & $75.7\%$ & $69.3\%$ & $93.3\%$ & $78.9\%$ & $61.9\%$ & $83.0\%$\\
    $2$ & $9$ & $86.4\%$ & $81.6\%$ & $79.2\%$ & $84.5\%$ & $74.1\%$ & $62.7\%$ & $77.6\%$ & $78.7\%$ & $55.5\%$ & $75.6\%$\\
    $2$ & $11$ & $93.1\%$ & $97.6\%$ & $98.7\%$ & $99.5\%$ & $92.3\%$ & $61.1\%$ & $96.8\%$ & $94.9\%$ & $83.5\%$ & $90.8\%$\\
    $2$ & $12$ & $97.1\%$ & $96.3\%$ & $84.8\%$ & $99.7\%$ & $79.7\%$ & $53.9\%$ & $91.2\%$ & $92.5\%$ & $61.9\%$ & $84.1\%$\\
    $2$ & $13$ & $95.5\%$ & $91.7\%$ & $81.1\%$ & $99.7\%$ & $76.5\%$ & $65.6\%$ & $92.8\%$ & $76.3\%$ & $64.8\%$ & $82.7\%$\\
    $2$ & $14$ & $95.5\%$ & $97.9\%$ & $97.9\%$ & $100.0\%$ & $76.5\%$ & $57.3\%$ & $95.5\%$ & $80.0\%$ & $83.5\%$ & $87.1\%$\\
    $2$ & $15$ & $90.9\%$ & $98.4\%$ & $84.3\%$ & $99.2\%$ & $89.6\%$ & $79.2\%$ & $88.3\%$ & $81.1\%$ & $72.5\%$ & $87.1\%$\\
    $2$ & $16$ & $93.6\%$ & $86.4\%$ & $70.1\%$ & $100.0\%$ & $75.2\%$ & $49.3\%$ & $75.5\%$ & $80.8\%$ & $54.4\%$ & $76.1\%$\\
    $2$ & $17$ & $94.4\%$ & $97.9\%$ & $92.0\%$ & $100.0\%$ & $78.7\%$ & $57.3\%$ & $83.5\%$ & $80.3\%$ & $60.0\%$ & $82.7\%$\\
    $2$ & $18$ & $96.5\%$ & $86.1\%$ & $80.3\%$ & $100.0\%$ & $70.7\%$ & $55.7\%$ & $80.0\%$ & $79.2\%$ & $64.3\%$ & $79.2\%$\\
    $2$ & $19$ & $92.5\%$ & $88.3\%$ & $88.5\%$ & $100.0\%$ & $74.9\%$ & $69.3\%$ & $89.3\%$ & $78.1\%$ & $59.7\%$ & $82.3\%$\\

    \hline
    \end{tabularx}
\end{table*}

\end{appendices}

\end{document}